\documentclass[authoryear]{elsarticle}

\usepackage{amsmath,amssymb}
\usepackage{graphicx}
\usepackage{hyperref}
\usepackage{wrapfig}
\usepackage{tikz}
\usepackage{bm}
\usepackage{subcaption} 
\usepackage{multirow}
\usepackage[table,xcdraw]{xcolor}
\usepackage{bigints}
\usepackage{setspace}
\usepackage{adjustbox}
\usepackage{xcolor} 
\usepackage[normalem]{ulem} 
\usepackage{cancel}
\usepackage[a4paper,margin=1in]{geometry}

\definecolor{LightRoyalBlue}{RGB}{163, 218, 255}
\definecolor{MediumRoyalBlue}{RGB}{100, 180, 255}
\definecolor{DarkRoyalBlue}{RGB}{4, 110, 202}

\newcommand{\Var}{\text{Var}}

\newcommand{\diff}{\text{d}}

\newcommand{\boldG}{\mathbf{G}}

\newcommand{\boldS}{\mathbf{S}}

\newcommand{\boldW}{\mathbf{W}}

\newcommand{\boldb}{\mathbf{b}}

\newcommand{\boldy}{\mathbf{y}}

\newcommand{\boldtheta}{\bm{\theta}}

\newcommand{\boldTheta}{\mathbf{\Theta}}

\newcommand{\textth}{$^{\text{th }}$}

\makeatletter
\def\ps@pprintTitle{%
  \let\@oddhead\@empty
  \let\@evenhead\@empty
  \let\@oddfoot\@empty
  \let\@evenfoot\@oddfoot}
\makeatother

\begin{document}
\doublespacing
\begin{frontmatter}

\title{Hierarchical Bayes meets hierarchical forecasting: A flexible framework for level-focused forecasts}

\author[aff1]{Arwen Nugteren\corref{cor1}}
\ead{a.nugteren@qut.edu.au}
\author[aff1]{Mahdi Abolghasemi}
\author[aff1]{Kerrie Mengersen}
\author[aff1]{Christopher Drovandi}

\cortext[cor1]{Corresponding author}

\affiliation[aff1]{organization={Queensland University of Technology, School of Mathematical Sciences, Centre for Data Science},
            addressline={2 George St},
            city={Brisbane},
            postcode={4000},
            country={Australia}}

\affiliation[aff2]{organization={Queensland University of Technology, Centre for Data Science},
            addressline={2 George St},
            city={Brisbane},
            postcode={4000},
            country={Australia}}

\affiliation[aff2]{organization={Second Affiliation},
            city={City},
            country={Country}}

\begin{abstract}
Decision-making in hierarchical systems requires probabilistic forecasts at all cross-sectional levels. Current hierarchical forecasting methods typically generate independent forecasts at each level and reconcile them post hoc to ensure coherence between upper and lower levels. Such post hoc corrections do not incorporate hierarchical structure or decision goals into the underlying parameter estimation. We propose a fully Bayesian hierarchical forecasting framework that shares information more effectively between and across levels than reconciliation alone. Our approach has the flexibility to softly penalise incoherence, subject to model specification, and to focus the global model and coherence update on hierarchical levels most relevant to decision outcomes. This yields parameter estimates that are focused towards the forecasting goals and capture the requirement for coherence, removing the need to estimate covariance matrices for multi-step forecasting horizons. We demonstrate improvements in predictive accuracy metrics on both simulated data and Australian domestic tourism forecasting. 
\end{abstract}


\end{frontmatter}


\section{Introduction}
Forecasting is rarely performed in isolation, but is instead typically used as the basis for decision-making. In large hierarchical systems, such as demand across different individual units, regions and totals, decision-makers require that forecasts at lower levels aggregate coherently to the forecasts at higher levels in order to make coherent decisions \citep{Athanasopoulos2024, Wen2026,Eckert2021}. They also require probabilistic forecasts to support appropriate planning in the face of uncertainty. Furthermore, while forecasting the entire hierarchy coherently tends to improve forecast quality at all levels \citep{Wickramasuriya2019,Panagiotelis2023,Athanasopoulos2024}, certain levels are likely to be more important to the decision outcomes \citep{Abolghasemi2022}, while strict coherency may not be required provided forecasts are sufficiently close to coherent to be useful in decision-making. Consequently, forecast and reconciliation methodologies should capture this differential importance. In this paper, we develop a Bayesian hierarchical approach to generate hierarchical probabilistic forecasts that share information between and across levels to improve forecast quality. We then apply a soft constraint to reduce incoherence across the forecast expected values, subject to the level of model misspecification. We combine this with the flexibility to weight levels on the basis of their decision importance, optimising the coherent hierarchy towards accuracy at a specific level of interest. 

Ensuring coherence of forecasts began with single-level (bottom-up, top-down and middle-out) approaches that guaranteed coherence, but did not leverage information at different levels \citep{Athanasopoulos2024}. Forecast reconciliation that involved post-processing base forecasts at all levels to ensure coherence of the means was first proposed by \citet{Hyndman2011} and later extended to the minimum trace (MinT) method of \citet{Wickramasuriya2019}, which has been proven to reduce the overall root mean squared error of reconciled forecasts. A review of forecast reconciliation, largely for point forecasts, is given by \citet{Athanasopoulos2024}. 

Probabilistic coherence of hierarchical forecasts has been a more recent focus in the literature. Some of the earliest examples are those of \citet{Jeon2019} employing quantile reconciliation in a temporal hierarchy and \citet{BenTaieb2021} using copula models on sub-hierarchies of parent-child nodes. A related approach is to model all the bottom-level series jointly, as done in a Bayesian framework by \citet{Bertani2025}. However, this becomes computationally challenging due to the very large number of series, possible covariates and modelling of the covariance structure. These approaches improve on single-level or independent models, but are limited in their applicability to a wider range of data and forecasting applications \citep{Panagiotelis2023}. 

\citet{Corani2021, Carrara2025,Corani2024} approach the problem of probabilistic forecast reconciliation in a post-processing step using Bayes rule, updating the base bottom-level series by conditioning on the base upper-level forecasts, then aggregating bottom-up. These ideas build on those in \citet{West1989}, which discuss using Bayes rule to condition on additional information or forecasts received for aggregate levels. In the Gaussian case of \citet{Corani2021}, the reconciliation-by-conditioning approach yields the same posterior means asymptotically as MinT reconciliation. The main limitation of this approach is that it is restricted to Gaussian or Student $t$ distributions in the closed-form update.\citet{Panagiotelis2023} and \citet{Girolimetto2024} advance probabilistic reconciliation by providing a clear definition of probabilistic coherence and extending MinT reconciliation to the probabilistic case, with the additional ability to train reconciliation weights to optimise a particular scoring rule. 

These recent works extend MinT reconciliation, which is the current state-of-the-art point forecast reconciliation method. However, MinT has several key limitations. Firstly, it is a post-processing step, and the coherency requirement is not reflected in the model parameter estimates. This limitation applies to all two-step reconciliation procedures. Secondly, the reconciliation requires estimation of the covariance matrix of base forecast errors, which is challenging for multi-step forecast horizons \citep{Girolimetto2024}. Thirdly, while it has been proven to decrease forecast error overall in the point forecasting case, this may be at the expense of performance at some levels \citep{Abolghasemi2022}, and no such theoretical result of optimality exists for the probabilistic case \citep{Panagiotelis2023}. It additionally does not allow the forecaster or decision-maker to control for relative importance or confidence in the forecasts at certain levels. 

Machine learning approaches to coherent hierarchical probabilistic forecasting have also seen some success \citep{Abolghasemi2019,Spiliotis2021}. \citet{Olivares2024b,Olivares2024} and \citet{Das2023} explore single-level approaches that are trained on the information from the full hierarchy, then aggregated or disaggregated to provide coherent probabilistic forecasts by construction. In these cases, explicit evaluation of the calibration at each level is not discussed. A more complete example is that of \citet{Rangapuram2021}, who develop a model where reconciliation is obtained by projection onto the coherent subspace in the final step of model training. While this is an end-to-end procedure, as the loss function used optimises accuracy with respect to the aggregation constraints, it does not return parameters of the coherent distribution, only reconciled samples. 

Most similar to our proposed approach is that of \citet{Kamarthi2023}, who also propose a soft probabilistic coherency constraint. They train a neural network to learn Gaussian parameters for their base forecasts, then apply a distributional coherency regularisation to update these parameters to a reconciled Gaussian distribution. This framework is extended to sparse time series in \citet{Kamarthi2024}, but is still limited to assuming the distribution is Poisson or Gaussian -- distributions that are closed under summation. 

In this work, we consider Bayesian hierarchical methods to perform hierarchical forecasting. Bayesian methods are well-suited to probabilistic forecasting, as the predictive distribution accounts for uncertainty in parameter estimates as well as the noise in the data. They have also been applied in hierarchical forecasting, leveraging informative priors to account for decisions or relative importance within a hierarchy \citep{Eckert2021}. A review of Bayesian forecasting approaches and their merit is given in \citet{Martin2024}. In particular, Bayesian hierarchical methods are a powerful tool for allowing partial information pooling informed by the data \citep{Gelman2013}. We leverage these properties to share information both between series at the same level and across hierarchical levels within a probabilistic framework. 

Bayesian hierarchical methods have been used in forecasting for a number of years. Early contributions include that of \citet{Gamerman1993}, providing an early example of bottom-up forecasting for Gaussian hierarchical models. They use a Bayesian hierarchical dynamic model for bottom-level series, then aggregate bottom-up following the approach of \citet{West1989} to obtain upper-level predictive distributions. 

\citet{Hollyman2022} and \citet{hollyman2024} build on the approach of \citet{West1997}, extending the framework to estimate the covariance matrix of all the series in the hierarchy using a hierarchical factor loadings model, addressing the challenging problem of estimating high-dimensional covariance matrices. This allows identification of which series are contributing the most to improvements in accuracy following reconciliation. These works are an interesting alternative to MinT reconciliation, but are still a two-step forecast-then-reconcile procedure.

Other applications of hierarchical Bayesian methods in forecasting include sales transaction forecasting \citep{Agosta2023}, building electricity demand \citep{Grillone2021}, residential electricity demand \citep{Wang2017} and the winner of the M6 competition forecasting track \citep{Weitzenfeld2025}. In these examples, the Bayesian hierarchical structure is used to improve parameter estimates through information pooling, but performance at aggregate levels or leverage of information available at upper levels is not considered. These works also do not evaluate the full probabilistic distribution to assess how the Bayesian hierarchical approach improves probabilistic forecasting.

Our contribution is developing a fully Bayesian framework for hierarchical forecasting that is able to focus on specific levels within a global forecasting model while softly adhering to aggregation constraints. In doing so, we demonstrate the value of Bayesian hierarchical methods for parameter estimation in hierarchical forecasting, improving information sharing both within and between hierarchical levels. The focusing approach integrates with the Bayesian hierarchical structure, so that the hierarchical model can be focused on a level with better data quality and hence improve forecasting performance at other levels. Our coherency penalisation approach is incorporated into the training of model parameters, allowing simple calculation of multi-horizon forecasts without the need to estimate multi-horizon covariance matrices of forecasting errors.

In this paper, we first discuss hierarchical forecasting and MinT reconciliation in Section \ref{sec:mint}, then present our own proposed methodology in Section \ref{methodology}. We apply our approach to simulated data to demonstrate the effect of each element of our methodology in Section \ref{simulation}, then demonstrate it on real data by forecasting Australian domestic tourism demand in Section \ref{tourism}, before providing a discussion in Section \ref{discussion} and concluding remarks in Section \ref{conclusion}.

\section{Hierarchical forecasting and reconciliation} \label{sec:mint}

This notation and terminology follow the guidelines of \citet{Hyndman2022}. We consider a simple hierarchy as shown in Figure \ref{fig:hierarchy_diagram}. A hierarchical time series over time $t = 1, \ldots T$ is a multivariate series $\boldy = \{\boldy_1, \ldots, \boldy_T\}$ where each $\boldy_t = (y_{t,1}, \ldots, y_{t,n})$ contains observations for all $n$ time series at different levels of the hierarchy. Given these data, we produce forecasts for a horizon $h$ steps ahead to estimate $\boldy_{T+h}$. We denote these forecasts as $\hat{\boldy}_{T+h \vert T}$, and refer to them as the base forecasts for $\boldy_{T+h}$.

We denote the most bottom-level $n_b$ series by $\boldb_t$. The time series can then be written as $\boldy_t = \boldS \boldb_t$, where  $\boldS$ is the structural matrix which encodes the linear constraints that must hold for this series to be coherent.

\begin{wrapfigure}{R}{0.4\textwidth}
\centering
\begin{tikzpicture}[level distance=2cm,
  level 1/.style={sibling distance=3cm},
  level 2/.style={sibling distance=1cm},
  every node/.style={circle, draw, minimum size=0.7cm}]
  \node[fill=MediumRoyalBlue] {Total}
    child {node[fill=LightRoyalBlue] {A}
      child {node {AA}}
      child {node {AB}}
    }
    child {node[fill=LightRoyalBlue] {B}
      child {node {BA}}
      child {node {BB}}
    };
\end{tikzpicture}
\caption{A simple three-level hierarchy.} \label{fig:hierarchy_diagram}
\end{wrapfigure}

All point linear reconciliation methods (including top-down and bottom-up) can be written as $\tilde{\boldy}_{T+h \vert T} = \boldS\boldG \hat{\boldy}_{T+h \vert T}$ for different choices of a reconciliation matrix $\boldG$, where $\tilde{\boldy}_{T+h \vert T}$ represents the reconciled forecast at time $T+h$. Sample-based probabilistic reconciliations can also be calculated in the same way, where $\hat{\boldy}_{T+h \vert T}$ is a single draw from the base distribution.

\subsection{Minimum trace reconciliation}
The current state-of-the-art method for point forecast reconciliation is minimum trace (MinT) reconciliation \citep{Wickramasuriya2019}. The method minimises the trace of the covariance matrix of the reconciled forecast errors, $\Var(\boldy_{T+h\vert T} - \tilde{\boldy}_{T+h})$. In Lemma 1 of their paper, they prove that this $h$-step-ahead covariance matrix of forecast reconciliation errors, conditional on the observed data up to time $T$, $\mathcal{I}_T$, can be expressed as, 
\begin{equation}
    \Var(\boldy_{T+h} - \tilde{\boldy}_{T+h} \vert \mathcal{I}_T) = \boldS \boldG \boldW_h \boldG' \boldS'
\end{equation}
for any $\boldG$ where $\boldS \boldG \boldS = \boldS$ and $\boldW_h$ is the covariance matrix of the $h$-step ahead base forecast errors. This relates the covariance of the reconciliation errors to the base forecast errors, which can be estimated from the data. The optimal reconciliation method is then given by choosing $\boldG$ as follows (Theorem 2 of \citet{Wickramasuriya2019}), 
\begin{equation}
    \boldG = (\boldS' \boldW_h^{-1} S)^{-1} \boldS' \boldW_h^{-1}.
\end{equation}

\citet{Wickramasuriya2019} prove that the MinT reconciliation method minimises the mean squared error of unbiased coherent forecasts, and that the MinT reconciled forecasts perform at least as well as the incoherent base forecasts. In practice, while mean squared error is improved overall, there is no guarantee of improvement at all levels, and worse performance may be observed at some levels. Estimating the covariance matrix, $\boldW_h$, for horizons $h>1$ is also a significant challenge \citep{Wickramasuriya2024}. \citet{Wickramasuriya2019} compare various approximations, typically expressing $\boldW_h$ as a scalar multiple of the identity matrix, $\boldW_1$ or a shrinkage estimator of $\boldW_1$, finding that shrinkage estimators typically show the best performance. \citet{DiFonzo2023} discuss other possible estimators for the covariance matrix in their extension of forecast reconciliation to the cross-temporal case, where a temporal hierarchy over different time intervals is introduced. However, the best-performing estimate was still a form of shrinkage estimator based on $\boldW_1$.

MinT reconciliation can also be extended to the probabilistic case by reconciling samples taken from the joint distribution over the hierarchy, or in closed form when the base forecast distribution is Gaussian \citep{Panagiotelis2023, Wickramasuriya2024}. Although this is an effective extension of the method to guarantee probabilistic coherence, theoretical guarantees of optimality with respect to the log score exist only in the case where the base forecast distribution is jointly Gaussian \citep{Wickramasuriya2024}.

\section{Methodology} \label{methodology}
Our approach is to generate univariate forecasts for each series, using a hierarchical Bayesian framework to learn parameters, apply a penalty on incoherence and allow for focusing on particular series of interest. This approach of generating univariate forecasts for each series, followed by ensuring coherence, is common in the hierarchical time series literature \citep{Hyndman2011, Wickramasuriya2019, Corani2021}, and exploits trends from all levels of the hierarchy while avoiding the need for a complex joint model. 

Our proposed methodology may be applied to any parametric likelihood, $p(\boldy \vert \boldtheta)$, with data $\boldy = \{\boldy_1, \ldots, \boldy_T\}$, where each $\boldy_t = (y_{t,1}, \ldots, y_{t,n})$ with corresponding parameters $\boldtheta = (\bm{\theta}_1, \ldots, \bm{\theta}_n)$. We train the model parameters using a Bayesian update, 

\begin{equation}
    p(\boldtheta \vert \boldy) = \frac{p(\boldy \vert \boldtheta) p(\boldtheta)}{p(\boldy)},
\end{equation}
where $p(\boldtheta \vert \boldy)$ is the posterior parameter distribution, $p(\boldy \vert \boldtheta)$ is the likelihood of the data given $\boldtheta$, $p(\boldy)$ is the overall probability of observing the data, and $p(\boldtheta)$ is the prior distribution on the parameters. We obtain one-step-ahead predictions, $\hat{\boldy}_{T+1}$, using the posterior predictive distribution, 
\begin{equation}
    p(\hat{\boldy}_{T+1} \vert \boldy) = \int_\boldTheta p(\hat{\boldy}_{T+1} \vert \boldtheta) p(\boldtheta \vert \boldy) \; \diff\boldtheta,
\end{equation}
where $\boldTheta$ is the sample space of the parameters. This posterior predictive distribution captures the uncertainty in both the unknown future prediction and the uncertainty in the parameter estimates, obtained by integrating with respect to the posterior \citep{Gelman2013}. Predictions for multiple steps ahead can be calculated by iteratively integrating over the predicted future values to capture the uncertainty in their predicted values.

\subsection{Bayesian hierarchical prior}
In a Bayesian model, priors range from informative distributions that captures prior knowledge or expert judgement to weakly informative priors that have little effect on the posterior \citep{Gelman2013}. We choose a hierarchical prior, in which we specify a hyperprior $p(\bm{\varphi})$ with hyperparameters $\bm{\varphi}$ for the parameters, $\bm{\theta}_i$ for the $i$\textth series within the hierarchy, $y_{1,i}, \ldots, y_{T,i}$. This captures the prior assumption that the parameters for each series are related. The prior distribution on the parameters associated with the $i$\textth series is then given by $p(\bm{\theta}_i \vert \bm{\varphi})$, yielding a joint prior for a given node, 

\begin{equation}
    p(\bm{\theta}, \bm{\varphi}) = p(\bm{\varphi}) \prod_{i=1}^n p(\bm{\theta}_i \vert \bm{\varphi}) .
\end{equation}

The use of the hyperprior, $p(\bm{\varphi})$, facilitates partial pooling of parameter estimates, shrinking extreme or uncertain estimates of $\boldtheta_i$ towards the mean parameter estimate for that node, but still allowing some individual variance between parameter estimates. This is particularly valuable for series with noisy or incomplete data, as they can be informed by related series. Where related series may differ significantly, it is appropriate to specify a larger hyperprior variance when choosing $p(\bm{\varphi})$.

\begin{figure}
\centering
\includegraphics[width=12cm]{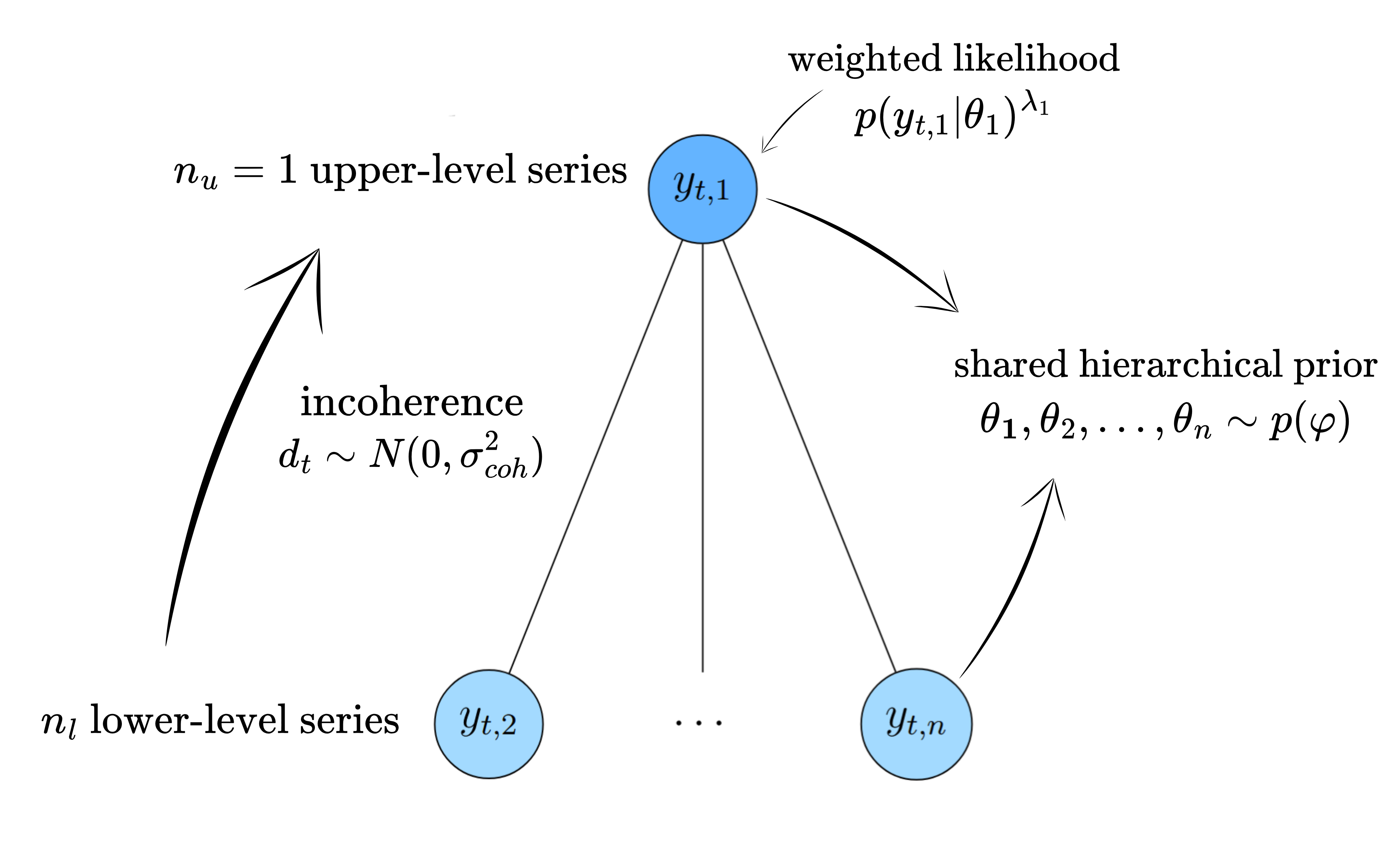}
\caption{A figure illustrating all three aspects of our methodology -- shared hierarchical prior, incoherence penalty and level-weighting -- on a simple hierarchy with an upper and lower level. Here, $\bm{\theta} = (\theta_1, \ldots, \theta_n)$. } \label{fig:methodology_diagram}
\end{figure}

We structure the parameter hierarchy to mirror the cross-sectional hierarchy that we are forecasting, such that at each node, the child and parent series share a hierarchical hyperprior, $p(\varphi)$, as illustrated for a two-level hierarchy in Figure \ref{fig:methodology_diagram}. In this way, the parameter hierarchy facilitates information transfer between series that share a parent node. If we have $m$ parent nodes and a third hierarchical level, we can extend the model by defining a hyperprior on $\bm{\varphi} = (\bm{\varphi}_1, \ldots, \bm{\varphi}_m) $ by $p(\bm{\varphi}\vert \bm{\xi}) =  p(\bm{\xi}) \prod\limits_{j=1}^m p(\bm{\varphi}_j \vert \bm{\xi})$. We also draw the original hyperparameters $\bm{\varphi}$, from $p(\bm{\xi})$. Additional hierarchical levels may be added by defining a further layer of hyperpriors. 

This sharing of the hyperprior between levels facilitates additional information sharing between levels where it is reasonable to assume series are not independent, and indirect information sharing at the lower levels from series further apart.

\subsection{Incoherence penalty}
We penalise incoherence between the fitted values of a set of $n_u$ upper-level series and $n_l$ lower-level series at time $t$,

\begin{equation}
    d_t =  \left(\sum_{i=1}^{n_u} \hat{\mu}_{t,i} - \sum_{i=n_u + 1}^{n_u + n_l} \hat{\mu}_{t,i}\right),
\end{equation}
as shown in Figure \ref{fig:methodology_diagram}. Note that this set of upper-level series could be a single parent node, or the set of all series at a particular level. 

We then penalise incoherence by multiplying the likelihood by $p\left(\bm{d}  \vert \bm{\theta}\right) = N(\bm{0}, \sigma_{\mathrm{coh}}^2 I_T)$, where $\bm{d} = (d_1, \ldots, d_T)$, $I_T$ is the $T \times T$ identity matrix and $\bm{\theta}$ are the time series parameters required to obtain the fitted values. The value of $\sigma_{\mathrm{coh}}$ is fixed and determines the level of penalisation applied to incoherence, which can be informed by decision-making requirements to maintain incoherence within a specified range. For example, using the fact that our penalty is Gaussian, 99\% of the incoherence values will fall within $\pm3\sigma_{\mathrm{coh}}$. 

Our use of a distributional constraint means that incoherence is only softly penalised, allowing the model to flexibly balance coherence against accuracy, avoiding rigid constraints that may produce unrealistic predictions, particularly in the presence of model misspecification. We choose this form of incoherence penalty such that both model fit and incoherence are penalised using the log-likelihood and scale similarly. More general loss functions could be used as a penalty, as in a generalised Bayes framework \citep{Bissiri2016}, but would require additional calibration parameters to ensure appropriate scaling.

Our posterior distribution for a two-level hierarchy (easily expanded to multiple levels) is then as follows: 

\begin{equation}
    p(\bm{\theta} \vert \boldy) \propto \underbrace{p(\bm{d} \vert \bm{\theta})   p(\boldy \vert \boldtheta) }_{\text{coherency-penalised likelihood}}  \underbrace{ \left( p(\bm{\varphi}) \prod_{i=1}^n p(\bm{\theta}_i \vert \bm{\varphi}) \right) }_{\text{hierarchical prior}}.
\end{equation}

We can extend the coherence penalty to an arbitrary number of levels by defining a new $\bm{d}$ and multiplying by a penalty $p(\bm{d} \vert \bm{\theta})$ for each pair of levels. The incoherence penalty can also be modified to handle grouped time series by penalising each group individually, again multiplying the penalties as above. In this case, we suggest setting $\sigma_\mathrm{coh}$ based on the scale of the data in the grouping, as we could reasonably expect different groups to have different magnitudes of raw data and incoherence.

\color{black}
\subsection{Level weighting}
It may be the case that a given series $\boldy_i$ or level is to be prioritised, as it is particularly relevant to decision-making or the data at that level is known to be more reliable. In such scenarios, we propose weighting this level by a factor $\lambda_i$, similar to a power posterior model \citep{Grunwald2012,McLatchie2025}, which use $\lambda < 1$ to decrease the weight of the data to improve robustness. Our proposal is to weight some series using $\lambda > 1$ to increase their relative importance or to apply a factor of $\lambda < 1$ to the levels to be down-weighted. Both strategies increase the weight of the focused level relative to the non-focused levels, but choices of $\lambda<1$ will result in the posterior being more informed by the prior, which is typically only appropriate when a suitable informative prior is chosen \citep{Grunwald2017}. 

After selecting a series $y_{1,i}, \ldots, y_{T,i}$ to focus the model, we then use the weighted likelihood given by $p(y_{1,i}, \ldots, y_{T,i} \vert \boldtheta_i)^{\lambda_i}$. If we are focusing on an entire level, we apply this penalty $\lambda_i$ to all series in the level. If implemented in combination with the hierarchical model, this greater emphasis on a particular level will carry through to update the parameters of the parent or child nodes, while maintaining the soft constraint on coherence with minimal adjustment to the weighted series, unlike a standard reconciliation approach. 

Our full posterior distribution is then of the form, 
\begin{equation}
    p(\bm{\theta} \vert \boldy) \propto \underbrace{ p(\bm{d} \vert \sigma_{\mathrm{coh}}^2 I_T) p(y_{1,1}, \ldots, y_{T,1} \vert \boldtheta_1)^{\lambda_1} \prod_{i=2}^{n} p(y_{1,i}, \ldots, y_{T,i} \vert \boldtheta_i)  }_{\text{weighted, coherency-penalised likelihood}} \underbrace{\left( p(\bm{\varphi}) \prod_{i=1}^n p(\bm{\theta}_i \vert \bm{\varphi}) \right) }_{\text{hierarchical prior}}, 
\end{equation}
where, for simplicity, we are weighting a single series $y_{1,1}, \ldots, y_{T,1}$ and writing a two-level hierarchical prior, as illustrated in Figure \ref{fig:methodology_diagram}. Our framework is flexible and can weight multiple series with different choices of weightings, or employ multiple levels of hierarchical priors to further share information. We are also able to weight coherency between particular levels or at specific nodes in a similar way to emphasise the relative importance of coherency at that location.

\subsection{Evaluation metrics for accuracy and coherency}
As our focus is on probabilistic forecasting, we use the unscaled energy score (ES) overall and the unscaled arithmetic mean of the continuous ranked probability score (CRPS) at each level to evaluate the probabilistic performance of our approaches. We avoid using log score as \citet{Panagiotelis2023} prove that it is not proper in the case of comparing coherent to incoherent hierarchies. We complement these probabilistic scoring rules by including the coverage of 95\% prediction intervals to assess calibration. To appropriately compare with MinT approaches, which have been proven to be optimal with respect to the mean \citep{Wickramasuriya2019}, we report both root mean squared scaled error (RMSSE) with respect to a naive or seasonal naive forecast as appropriate for overall accuracy and the average root mean squared error (RMSE) at each level. 

For assessing coherency, we define mean absolute incoherence (MAI) between some set of $n_u$ upper-level series and $n_l$ lower-level series, 

\begin{equation}
    \mbox{MAI}\left(\sum_{i=1}^{n_u} y_{t,i}, \sum_{i=n_u + 1}^{n_u + n_l} y_{t,i}\right) = \left\vert \sum_{i=1}^{n_u} y_{t,i} - \sum_{i=n_u + 1}^{n_u + n_l} y_{t,i} \right\vert. 
\end{equation}
In our results, we report the global incoherence between consecutive levels and between the top-most and bottom-most levels.

\section{Simulation study} \label{simulation}
We first test our proposed methodology on simulated data to demonstrate its behaviour in scenarios where there is limited training data, varying levels of misspecification and situations where the data itself is incoherent. We generate synthetic data from a similar setup to that of \citet{Wickramasuriya2019} and \citet{Corani2021}, generating bottom-level ARMA(2,2) processes with correlated Gaussian error following the 3-level hierarchy in Figure \ref{fig:hierarchy_diagram}, and evaluate forecasting horizons $h=1,\ldots,5$ over a 500-window rolling origin. For our model, we intentionally introduce misspecification, as knowledge of the true data-generating process is not realistic in practical forecasting applications. To do this, we employ an ETS model to forecast the mean, with Laplace-distributed independent errors. This choice means that the model does not capture correlation between series, autocorrelation in the error processes, and Laplace distributions do not aggregate to another Laplace distribution. 

We test both independent Bayesian and hierarchical Bayesian models, and apply the MinT (sample) method of \citet{Panagiotelis2023} for probabilistic forecast reconciliation on both as a benchmark for comparison. These are reported as independent Bayes MinT and hierarchical Bayes MinT. \color{black} For the coherency penalty, we set $\sigma_{\mathrm{coh}} = 1$ and $\sigma_{\mathrm{coh}} = \sqrt{0.5}$ for the softer and tighter coherency penalties respectively. These values correspond to 99\% of incoherence being controlled within 35\% and 20\% of the average mean absolute error of the top level in the well-specified independent Bayesian model. Tighter penalties are not appropriate due to the relatively large variance in the simulated data. We test two weighting penalties: weighting the top level by the number of bottom-level series feeding into it (Hierarchical Bayes total-weighted), and weighting the bottom level by a factor of 2 (Hierarchical Bayes bottom-weighted), representing a doubling of the weight of this data. We choose these penalties as initial exploration of the data revealed that the misspecified model is less misspecified at the bottom level compared to the top level, so we would expect better overall performance from weighting the bottom levels compared to the top. For both focused models, we use the tighter coherency penalty of $\sigma_{\mathrm{coh}} = \sqrt{0.5}$. 

Complete details of the simulation setup and statistical models, and comparative results for a well-specified data generating process are included in \ref{appendix:simulation}.

\subsection{Results}

Table \ref{tab:simulation_windowsize} shows the value of employing a Bayesian hierarchical structure over independent Bayesian estimation in small sample sizes, as the training window size varies between 10, 20 and 50. The percentage improvement of the Bayesian hierarchy over the independent Bayesian method shows that the Bayesian hierarchy provides the most value when the available training data is limited, with the benefit decreasing as the training data size increases. This is expected, as when data is limited for an individual series, parameters are learnt by borrowing information from neighbouring series. However, when more data is available, there is sufficient information in the data to learn optimal parameters in each series. All following results are reported for a training window size of 50.

\begin{table}[h!] 
\caption{Results for the overall improvement of the Bayesian hierarchy over independent Bayes methods for training window sizes of 10, 20 and 50. All methods had the same evaluation set and have been compared against the same naive forecast for computation of RMSSE values. Percentage differences were calculated prior to rounding.}
\label{tab:simulation_windowsize}
\begin{adjustbox}{max width=\textwidth}
\begin{tabular}{lllll}
Training data size   & \multicolumn{1}{l|}{Method}                 & RMSSE         & ES            & 95\% Coverage \\ \hline
\rowcolor[HTML]{EFEFEF} 
                     &                                             &               &               &               \\
10                   & \multicolumn{1}{l|}{Independent Bayes}      & 0.82          & 16.8          & 98.4          \\
                     & \multicolumn{1}{l|}{Hierarchical Bayes}     & \textbf{0.79} & \textbf{16.0} & \textbf{97.3} \\
                     & \multicolumn{1}{l|}{Percentage improvement} & 3.82\%        & 4.75\%        & 1.07\%        \\
                     &                                             &               &               &               \\
                     & \multicolumn{1}{l|}{Independent Bayes}      & 0.75          & 15.0          & 98.5          \\
                     & \multicolumn{1}{l|}{Hierarchical Bayes}     & \textbf{0.74} & \textbf{14.7} & \textbf{97.6} \\
\multirow{-3}{*}{20} & \multicolumn{1}{l|}{Percentage improvement} & 1.13\%        & 1.55\%        & 0.84\%        \\
                     &                                             &               &               &               \\
                     & \multicolumn{1}{l|}{Independent Bayes}      & \textbf{0.70} & 14.0          & 97.7          \\
                     & \multicolumn{1}{l|}{Hierarchical Bayes}     & \textbf{0.70} & \textbf{13.9} & \textbf{97.5} \\
\multirow{-3}{*}{50} & \multicolumn{1}{l|}{Percentage improvement} & 0.24\%        & 0.45\%        & 0.37\%       
\end{tabular}
\end{adjustbox}
\end{table}

Table \ref{tab:simulation_coherence} shows a clear pattern of incoherence decreasing as penalisation is applied. The Bayesian hierarchical methods have significantly lower mean incoherence between all levels, even before applying any incoherence penalisation. When we penalise incoherence, we observe that the mean incoherence drops significantly, even when the incoherence is softly penalised.

\begin{table}[h!]
\centering
\caption{Mean absolute incoherence between pairs of levels, excluding MinT reconciliation methods, which have no incoherence by construction.} \label{tab:simulation_coherence}
\footnotesize
\begin{adjustbox}{max width=\textwidth}
\begin{tabular}{llll}
Method                                                    & \begin{tabular}[c]{@{}l@{}}Top:\\ Bottom\end{tabular} & \begin{tabular}[c]{@{}l@{}}Top:\\ Middle\end{tabular} & \begin{tabular}[c]{@{}l@{}}Middle:\\ Bottom\end{tabular}                     \\ \hline
\rowcolor[HTML]{EFEFEF} 
                                                          &                                                       &                                                       &                                                                              \\
\multicolumn{1}{l|}{Independent Bayes}                    & 2.09                                                  & 1.69                                                  & 2.01                                                                         \\
\multicolumn{1}{l|}{Hierarchical Bayes}                   & 1.74    & 1.28    & 1.48                          \\
\multicolumn{1}{l|}{Hierarchical Bayes softer coherency}  & 0.21    & 0.16    & 0.19                                                                         \\
\multicolumn{1}{l|}{Hierarchical Bayes tighter coherency} & \textbf{0.18}                        & \textbf{0.15}                        & \textbf{0.16} \\
\multicolumn{1}{l|}{Hierarchical Bayes total-weighted}    & 0.22                                                  & 0.17                                                  & 0.18                          \\
\multicolumn{1}{l|}{Hierarchical Bayes bottom-weighted}   & 0.20                                                  & 0.15                                                  & 0.18                                                                        
\end{tabular}
\end{adjustbox}
\end{table}

\color{black}
Table \ref{tab:simulation_accuracy} shows there is minimal overall difference between the methods, with all of them performing well at different levels, and no differences in RMSSE to two decimal places at the top level. MinT methods show better overall energy scores and RMSE, in line with their theoretical properties, but undercover. 

Focusing on the top level has improved top-level results relative to the other Bayes coherency approaches, but has resulted in worse overall performance due to the focusing approach. Weighting the bottom level has been beneficial, improving RMSE and CRPS at the bottom level and overall, due to the relatively lower degree of misspecification at the bottom level. Specifically, we compare log Bayes factors at each level relative to a well-specified model \citep{Mitchell2011}, where we see that the bottom-level independent model has a log Bayes factor of 0.19, while this increases to a log Bayes factor of 0.36 at the top level, indicating an increase in model misspecification. It is clear from the results that while we obtain level-specific performance improvement, focusing on highly misspecified levels can decrease overall performance, while focusing on better-specified levels can improve it. 

\begin{table}[h!]
\caption{Overall accuracy results at all levels.} \label{tab:simulation_accuracy}
\footnotesize
\begin{tabular}{lllll}
Level                     & Method                                                    & RMSSE                                               & ES                                                 & 95\% Coverage                                                                \\ \hline
\rowcolor[HTML]{EFEFEF} 
                          &                                                           &                                                     &                                                    &                                                                              \\
                          & \multicolumn{1}{l|}{Independent Bayes}                    & \textbf{0.70}                      & 14.0                                               & 97.7                                                                         \\
                          & \multicolumn{1}{l|}{Independent Bayes MinT}               & \textbf{0.70}                      & \textbf{13.7}                     & 91.0                                                                         \\
                          & \multicolumn{1}{l|}{Hierarchical Bayes}                   & \textbf{0.70}                      & 13.9                                               & 97.5                          \\
                          & \multicolumn{1}{l|}{Hierarchical Bayes MinT}              & \textbf{0.70}                      & \textbf{13.7}                     & 90.3                           \\
                          & \multicolumn{1}{l|}{Hierarchical Bayes softer coherency}  & \textbf{0.70}                      & 13.9                                               & 97.1                                                                         \\
                          & \multicolumn{1}{l|}{Hierarchical Bayes tighter coherency} & \textbf{0.70}                      & 13.9                                               & 97.1                                                                         \\
                          & \multicolumn{1}{l|}{Hierarchical Bayes total-weighted}    & \textbf{0.70}                      & 14.0                                               & \textbf{96.9}                                               \\
\multirow{-8}{*}{Overall} & \multicolumn{1}{l|}{Hierarchical Bayes bottom-weighted}   & \textbf{0.70}                      & 13.8                                               & \textbf{96.9}                                               \\
                          &                                                           &                                                     &                                                    &                                                                              \\
Level                     & Method                                                    & RMSE                                                & CRPS                                               & 95\% Coverage                                                                \\ \hline
\rowcolor[HTML]{EFEFEF} 
                          &                                                           &                                                     &                                                    &                                                                              \\
                          & \multicolumn{1}{l|}{Independent Bayes}                    & \textbf{12.50}                     & \textbf{7.15}                     & 98.0                                                                         \\
                          & \multicolumn{1}{l|}{Independent Bayes MinT}               & 12.52                                               & 7.40                                               & 80.5                                                                         \\
                          & \multicolumn{1}{l|}{Hierarchical Bayes}                   & 12.61 & 7.21 & 97.2                           \\
                          & \multicolumn{1}{l|}{Hierarchical Bayes MinT}              & 12.61 & 7.49 & 79.4                           \\
                          & \multicolumn{1}{l|}{Hierarchical Bayes softer coherency}  & 12.84 & 7.31                                               & 96.0                          \\
                          & \multicolumn{1}{l|}{Hierarchical Bayes tighter coherency} & 12.85 & 7.31                                               & 95.9                           \\
                          & \multicolumn{1}{l|}{Hierarchical Bayes total-weighted}    & 12.94 & 7.40                                               & \textbf{95.4}                                               \\
\multirow{-8}{*}{Top}     & \multicolumn{1}{l|}{Hierarchical Bayes bottom-weighted}   & 12.77 & 7.26                                               & 96.1                          \\
\rowcolor[HTML]{EFEFEF} 
                          &                                                           &                                                     &                                                    &                                                                              \\
                          & \multicolumn{1}{l|}{Independent Bayes}                    & 8.37                                                & 4.79                                               & 97.5                                                                         \\
                          & \multicolumn{1}{l|}{Independent Bayes MinT}               & \textbf{8.26}                      & 4.79                                               & 87.3                                                                         \\
                          & \multicolumn{1}{l|}{Hierarchical Bayes}                   & 8.32  & \textbf{4.77}                     & 97.4                           \\
                          & \multicolumn{1}{l|}{Hierarchical Bayes MinT}              & 8.28                                                & 4.81 & 86.2                           \\
                          & \multicolumn{1}{l|}{Hierarchical Bayes softer coherency}  & 8.37                                                & 4.79                                               & 97.0                                                                         \\
                          & \multicolumn{1}{l|}{Hierarchical Bayes tighter coherency} & 8.38                                                & 4.80                                               & 96.9                           \\
                          & \multicolumn{1}{l|}{Hierarchical Bayes total-weighted}    & 8.38                                                & 4.80                                               & \textbf{96.7} \\
\multirow{-8}{*}{Middle}  & \multicolumn{1}{l|}{Hierarchical Bayes bottom-weighted}   & 8.33  & \textbf{4.77}                     & 97.1                                                                         \\
\rowcolor[HTML]{EFEFEF} 
                          &                                                           &                                                     &                                                    &                                                                              \\
                          & \multicolumn{1}{l|}{Independent Bayes}                    & 5.86                                                & 3.35                                               & 97.8                                                                         \\
                          & \multicolumn{1}{l|}{Independent Bayes MinT}               & 5.84                                                & 3.33                                               & 95.5                                                                         \\
                          & \multicolumn{1}{l|}{Hierarchical Bayes}                   & 5.84                                                & 3.34                                               & 97.6                           \\
                          & \multicolumn{1}{l|}{Hierarchical Bayes MinT}              & 5.82                                                & 3.33                                               & \textbf{95.1} \\
                          & \multicolumn{1}{l|}{Hierarchical Bayes softer coherency}  & 5.81                                                & \textbf{3.32}                     & 97.5                                                                         \\
                          & \multicolumn{1}{l|}{Hierarchical Bayes tighter coherency} & 5.81  & 3.33                                               & 97.5                          \\
                          & \multicolumn{1}{l|}{Hierarchical Bayes total-weighted}    & 5.82                                                & 3.33                                               & 97.3                          \\
\multirow{-8}{*}{Bottom}  & \multicolumn{1}{l|}{Hierarchical Bayes bottom-weighted}   & \textbf{5.79}                      & \textbf{3.32}                     & 97.0                                                                        
\end{tabular}
\end{table}

\subsection{Softly coherent data experiment}
We also assess the effect of our soft incoherence penalty when the data itself is softly incoherent. Such scenarios may arise in practice when hierarchical data is measured at each level, introducing measurement noise or bias. To model such a scenario, we generate incoherent data from the original simulated data by adding Gaussian noise independently to each series, where the standard deviation of the noise is set such that 99\% of the simulated noise falls within 5\% of the range of that series. In this experiment, we include comparisons to MinT methods, as well as the corresponding baseline independent and hierarchical Bayesian methods, to assess the relative performance of our soft coherency penalty against both entirely incoherent forecasts and perfectly coherent forecasts. When calculating RMSSE, we use the naive forecast based on the respective dataset, coherent and softly coherent (the trend of best overall RMSSE performance for our soft coherence penalty is consistent when compared against either coherent or incoherent naive forecasts).  

\begin{table}[h!]
\caption{Comparison of overall accuracy results between the model run on coherent and incoherent datasets. } \label{tab:incoherent_data}
\begin{adjustbox}{max width=\textwidth}
\begin{tabular}{llllllll}
                                                          & \multicolumn{3}{l}{Coherent data}             &  & \multicolumn{3}{l}{Incoherent data}           \\
\multicolumn{1}{l|}{Method}                               & RMSSE         & ES            & 95\% Coverage &  & RMSSE         & ES            & 95\% Coverage \\ \hline
\rowcolor[HTML]{EFEFEF} 
                                                          &               &               &               &  &               &               &               \\
\multicolumn{1}{l|}{Independent Bayes}                    & \textbf{0.70} & 14.0          & 97.7          &  & 0.70          & 14.1          & 97.8          \\
\multicolumn{1}{l|}{Independent Bayes MinT}               & \textbf{0.70} & \textbf{13.7} & 91.0          &  & 0.69          & \textbf{13.7} & 90.9          \\
\multicolumn{1}{l|}{Hierarchical Bayes}                   & \textbf{0.70} & 13.9          & 97.5          &  & 0.70          & 14.0          & 97.5          \\
\multicolumn{1}{l|}{Hierarchical Bayes MinT}              & \textbf{0.70} & \textbf{13.7} & 90.3          &  & 0.70          & 13.8          & 90.2          \\
\multicolumn{1}{l|}{Hierarchical Bayes softer coherency}  & \textbf{0.70} & 13.9          & \textbf{97.1} &  & 0.69          & 13.8          & \textbf{97.2} \\
\multicolumn{1}{l|}{Hierarchical Bayes tighter coherency} & \textbf{0.70} & 13.9          & \textbf{97.1} &  & \textbf{0.68} & \textbf{13.7} & 97.3         
\end{tabular}
\end{adjustbox}
\end{table}

We can see in Table \ref{tab:incoherent_data} that on the incoherent data, both our RMSSE and ES improve for our coherency penalised models relative to their performance on the coherent data. The results are also better for both our coherency penalised models over the independent and Bayesian hierarchical baselines on the incoherent data. They also performed equally well or better than the MinT methods, despite comparatively worse performance on the coherent data. This indicates that our approach of softly penalising incoherence may be effective when the data itself has measurement error obscuring the trend.

\begin{table}[h!]
\caption{Mean absolute incoherence for the misspecified model on both the coherent and softly incoherent data, compared to the actual incoherence in the data.} \label{tab:incoherent_data_MAI}
\begin{adjustbox}{max width=\textwidth}
\begin{tabular}{llllllll}
                                                          & \multicolumn{3}{l}{Coherent data}       &  & \multicolumn{3}{l}{Incoherent data}           \\
\multicolumn{1}{l|}{Method}                               & Top:Bottom & Top:Middle & Middle:Bottom &  & Top:Bottom    & Top:Middle    & Middle:Bottom \\ \hline
\rowcolor[HTML]{EFEFEF} 
Raw data                                                  & -          & -          & -             &  & 1.16          & 1.32          & 1.27          \\
\multicolumn{1}{l|}{Independent Bayes}                    & 2.09       & 1.69       & 2.01          &  & 2.27          & 1.93          & 2.37          \\
\multicolumn{1}{l|}{Hierarchical Bayes}                   & 1.74       & 1.28       & 1.48          &  & 1.90          & 1.54          & 1.72          \\
\multicolumn{1}{l|}{Hierarchical Bayes softer coherency}  & 0.21       & 0.16       & 0.19          &  & 0.30          & 0.25          & 0.26          \\
\multicolumn{1}{l|}{Hierarchical Bayes tighter coherency} &\textbf{0.18} & \textbf{0.15} & \textbf{0.16} &  & \textbf{0.22} & \textbf{0.20} & \textbf{0.20}
\end{tabular}
\end{adjustbox}
\end{table}

Table \ref{tab:incoherent_data_MAI} shows the MAI for the coherent and softly incoherent datasets. First, we observe that the incoherence in the raw simulated data is on a similar scale but slightly less than the incoherence in the non-penalised forecasts. We also see that for all 4 models compared, there is slightly higher incoherence for the forecasts trained on incoherent data; however, the overall trend of decreasing incoherence with increasing incoherence penalisation remains. It is also interesting to compare with Table \ref{tab:incoherent_data} and see that although there is slightly higher incoherence, overall accuracy has improved. This further suggests our approach may provide some benefit in this type of scenario, where incoherence is present in the actual data.

\color{black}
\section{Empirical evaluation: Forecasting Australian domestic tourism} \label{tourism}
We test our modelling framework on the Australian domestic tourism dataset of \citet{Wickramasuriya2019} used in testing hierarchical forecasting methodology by \citet{Wickramasuriya2019,Corani2021,Girolimetto2024,Carrara2025,Rangapuram2021,Bertani2025} and \citet{Kamarthi2023}, among others. This dataset consists of the number of nights spent away from home by Australians monthly from January 1998 to December 2016, split into 4 geographical levels: 76 regions, organised into 27 zones, 7 states/territories and the aggregate total for Australia. These are further divided into 4 purposes of travel, yielding a grouped hierarchy. We choose to model it as a single 5-level hierarchy, forecasting tourism flows split up by purpose of travel at each of the four geographical levels, then a fifth overall level aggregating the purposes of travel. We train the model on a rolling window of 8 years (96 observations) of historical data, and produce forecasts for 1-12 months ahead at each window. Additional details about this dataset are in \citet{Wickramasuriya2019}.

In all cases, our base forecasting model is an exponential smoothing (ETS) model, due to its success in modelling this dataset in \citet{Wickramasuriya2019}. Bayesian ETS is also established as an effective time series forecasting method \citep{Smyl2025}. Preliminary experiments showed little difference between additive and multiplicative ETS models for the error process, so we model all series with an additive ETS model. The error process was chosen to be Gaussian, in line with the most common approach in the other papers benchmarking against this dataset. This gives the following model structure for each individual time series: 

\begin{align}
    y_{t+h} & = \ell_t + s_{t+h-m(k+1)} + \varepsilon_t \\
    \ell_t &= \alpha(y_t - s_{t-m}) + (1-\alpha) \ell_{t-1}  \\
    s_t & = \gamma(y_t - \ell_{t-1}) + (1-\gamma)s_{t-m} \\
    \varepsilon_t &\sim N(0, \sigma^2),
\end{align}
where $t$ is the current timestep, $h$ is the forecasting horizon, $m$ is the seasonal period, $k$ is the integer part of $(h-1)/m$, $0 \leq \alpha \leq 1$ is the smoothing parameter for level, and $0 \leq \gamma \leq 1-\alpha$ is the smoothing parameter associated with seasonality. We model a yearly seasonality, so $m=12$. We need to estimate the coefficients $\alpha$ and $\gamma$, as well as initial seasonal values $\ell_0$ and $s_1, \ldots, s_{12}$, and the variance parameter $\sigma$. 

In our experiments, we use an independent Bayesian model as our baseline, where the parameters for each series are estimated independently. We then compare this to a hierarchical Bayesian model, where the parameter hierarchy follows the geographical hierarchical structure, a softer and tighter coherency-penalised hierarchical Bayesian models, and two level-weighted coherency-penalised hierarchical Bayesian models weighting the top level and the purpose-of-travel level. We also include MinT(Shrink) sample-based probabilistic reconciled versions \citep{Panagiotelis2023} of the independent Bayesian and hierarchical Bayesian models to compare their performance (independent Bayes MinT and hierarchical Bayes MinT respectively). We use the shrinkage estimator of the covariance matrix and use the approximation $\widehat{W}_h = k_h \widehat{W_1}$ with $k_h=1$, where the residuals are calculated using the posterior mean fitted values for the 1-step ahead forecasts.

\subsection{Data processing and software details}
All models are trained and run in Stan, which performs MCMC sampling using a Metropolis-Hastings algorithm \citep{Carpenter2017}. Stan models tend to perform better with standardised data, so we standardised the bottom-level series by the overall mean and standard deviation of the bottom level. In order to preserve the aggregation structure, we scaled the aggregate series by the bottom-level standard deviation, and the bottom-level mean multiplied by the number of bottom-level series feeding into the parent series.  

The model is run in Stan using 10 chains, with 2000 samples each, 1000 of which are discarded as burn-in, for a total of 10,000 posterior samples. Effective sample sizes were above 1500. For the independent and Bayesian hierarchical trials, model fit and sampling took under 2 hours per window. For the coherency-penalised trials, the tighter coherency model took up to 4 hours per window, due to the additional posterior complexity induced by requiring parameters to softly satisfy coherence. All experiments were run on a high-performance computing system.

\subsection{Priors}
Given the wide variability in tourism numbers between different regions, the initial values, $\ell_0$ and $s_1, \ldots, s_{12}$, are not modelled hierarchically in any of our experiments. At the bottom level, $\ell_0$ and $s_1, \ldots, s_{11}$ are given $N(0,1)$ priors; at aggregate levels they are given as $N(0, n_{b_i}^2)$ priors, where $n_{b_i}$ represents the number of bottom-level series feeding into that node. These priors are chosen to be weakly informative, in that we would expect most values to fall within the range $[0,n_{b_i}]$, so reasonable prior probability is assigned to all values within this range, but weakly enough that the data guides concentration onto a posterior value. As the seasonal coefficients should sum to 1, $s_{12}$ is estimated deterministically as $s_{12} = 1 - \sum_{i=1}^{11} s_i$ for each series. 

Variance parameters are estimated using half-Cauchy priors, as recommended in the Stan reference manual for Gaussian distributions \citep{stan_users_guide_2026}, using $\text{half-Cauchy}(0, 0.3)$ for the two lowest levels, $\text{half-Cauchy}(0, 0.5)$ for the next two middle levels, and $\text{half-Cauchy}(0,1)$ for the top level, representing weakly informative priors on the scaled data at each level.

For our baseline independent Bayesian model, priors for $\alpha$ and $\gamma$ are chosen to be $U(0,1)$, with the restriction $0 \leq \gamma \leq 1 - \alpha$. Diagnostic plots of the posteriors were assessed, and concentration away from the original uniform prior was observed.

In the hierarchical Bayesian trials, the $\alpha$ and $\gamma$ parameters controlling level and seasonal change over time are estimated in a Bayesian hierarchical structure, where the Bayesian hierarchy over the parameters follows the geographical hierarchy. Let $\bm{\theta} = (\alpha, \gamma)$ be parameters of a given series, and $\bm{\tau} = (\mu_\alpha, \mu_\gamma)$ be the hyperparameters of the same series. Then, at the very top level, we draw the hyperparameters from a top-level hyperprior: 
\begin{align}
    \mu_{\alpha} &\sim U(0,1), \\
    \mu_\gamma &\sim U(0,1-\mu_\alpha).
\end{align}

From this top-level hyperprior, we then draw hyperprior parameters at each level: $\bm{\tau}_{\mathrm{purposes}}$, $\bm{\tau}_{\mathrm{states}}$, $\bm{\tau}_{\mathrm{zones}}$ and $\bm{\tau}_{\mathrm{regions}}$ from Gaussian distributions centred around the mean of the level above with a relatively tight standard deviation of 0.1. This choice of standard deviation performed well in modelling for this dataset, but a wider hyperprior variance should be selected if the groups are known to differ significantly. Hyperprior distributions other than Gaussian can also be chosen if more appropriate for the structure of the system. 

Individual series parameters are then drawn from the hyperprior at that level, again modelled as normal around the hyperprior mean of that level, with a standard deviation of 0.1, as follows for each series $i$ in a given level: 
\begin{align}
    \alpha_{i, \mathrm{level}} &\sim N(\mu_{\alpha, \mathrm{level}}, 0.1^2), \\
    \gamma_{i, \mathrm{level}} &\sim N(\mu_{\gamma, \mathrm{level}}, 0.1^2).
\end{align}

Prior predictive checks were performed on a representative series at each hierarchical level, which verified the weakly informative nature of the selected priors. See Figures \ref{fig:prior_predictive_tourism_independent} and \ref{fig:prior_predictive_tourism_hierarchical} in the supplementary material for details.

\subsection{Coherency and level-weighting}
Coherency is penalised on the scaled data using standard deviations of $\sigma_{\mathrm{coh}} = 0.005 \cdot n_{b}$ and $\sigma_{\mathrm{coh}} = 0.002 \cdot n_{b}$  for the softer and tighter penalties, respectively, where $n_b$ is the number of bottom-level series. These correspond to 99\% incoherence values falling within 50\% and 20\% respectively of the average mean absolute error at the top level of the independent Bayesian model, ensuring that the incoherence error is much smaller than the forecast error. Tighter coherency penalties were also tested, but the model did not converge, indicating a lack of a significantly more coherent solution. We apply this penalty globally across levels, rather than individually at each node. 

For the level-weighted models, we choose to focus on the upper-level aggregate series, demonstrating the effect of weighting both the overall total and the total in each purpose by the average number of bottom-level series feeding into each. We apply the tighter coherency penalty of $\sigma_{\mathrm{coh}} = 0.002 \cdot n_b$ to each of these.

\subsection{Results}

\begin{table}[h!]
\begin{center}
\caption{Summary of the key accuracy metrics and coverage of 95\% prediction intervals for the competing methods at each level, averaged over all horizons.} \label{tab:overall_results}
\footnotesize
\begin{tabular}{lllll}
                            & Method                                                    & RMSSE          & ES            & 95\% Coverage \\ \hline
\rowcolor[HTML]{EFEFEF} 
                            &                                                           &                &               &               \\
                            & \multicolumn{1}{l|}{Independent Bayes}                    & 0.865          & 1985          & 95.4          \\
                            & \multicolumn{1}{l|}{Independent Bayes MinT}               & 0.869          & 1957          & 96.1          \\
                            & \multicolumn{1}{l|}{Hierarchical Bayes}                   & 0.858          & 1971          & 95.4          \\
                            & \multicolumn{1}{l|}{Hierarchical Bayes MinT}              & 0.860          & 1953          & 94.6          \\
                            & \multicolumn{1}{l|}{Hierarchical Bayes softer coherency}  & 0.858          & 1998          & \textbf{95.3} \\
                            & \multicolumn{1}{l|}{Hierarchical Bayes tighter coherency} & 0.858          & 2012          & \textbf{95.3} \\
                            & \multicolumn{1}{l|}{Hierarchical Bayes total-weighted}    & \textbf{0.857} & \textbf{1947} & \textbf{95.3} \\
\multirow{-8}{*}{Overall}   & \multicolumn{1}{l|}{Hierarchical Bayes purposes-weighted} & \textbf{0.857} & \textbf{1947} & \textbf{95.3} \\
                            &                                                           &                &               &               \\
Level                       & Method                                                    & RMSE           & CRPS          & 95\% Coverage \\ \hline
\rowcolor[HTML]{EFEFEF} 
                            &                                                           &                &               &               \\
                            & \multicolumn{1}{l|}{Independent Bayes}                    & 1817           & 1037          & \textbf{94.7} \\
                            & \multicolumn{1}{l|}{Independent Bayes MinT}               & 1782           & 1034          & 88.2          \\
                            & \multicolumn{1}{l|}{Hierarchical Bayes}                   & 1806           & 1030          & 94.3          \\
                            & \multicolumn{1}{l|}{Hierarchical Bayes MinT}              & 1782           & 1054          & 79.5          \\
                            & \multicolumn{1}{l|}{Hierarchical Bayes softer coherency}  & 1861           & 1072          & 90.8          \\
                            & \multicolumn{1}{l|}{Hierarchical Bayes tighter coherency} & 1884           & 1088          & 90.6          \\
                            & \multicolumn{1}{l|}{Hierarchical Bayes total-weighted}    & \textbf{1764}  & \textbf{1010} & 90.0          \\
\multirow{-8}{*}{Australia} & \multicolumn{1}{l|}{Hierarchical Bayes purposes-weighted} & 1785           & 1023          & 90.1          \\
\rowcolor[HTML]{EFEFEF} 
                            &                                                           &                &               &               \\
                            & \multicolumn{1}{l|}{Independent Bayes}                    & 688            & 381           & \textbf{93.3} \\
                            & \multicolumn{1}{l|}{Independent Bayes MinT}               & 686            & 381           & 90.7          \\
                            & \multicolumn{1}{l|}{Hierarchical Bayes}                   & 687            & 379           & 92.8          \\
                            & \multicolumn{1}{l|}{Hierarchical Bayes MinT}              & 683            & 381           & 86.0          \\
                            & \multicolumn{1}{l|}{Hierarchical Bayes softer coherency}  & 700            & 386           & 91.5          \\
                            & \multicolumn{1}{l|}{Hierarchical Bayes tighter coherency} & 707            & 391           & 91.1          \\
                            & \multicolumn{1}{l|}{Hierarchical Bayes total-weighted}    & 682            & 376           & 92.0          \\
\multirow{-8}{*}{Purposes}  & \multicolumn{1}{l|}{Hierarchical Bayes purposes-weighted} & \textbf{675}   & \textbf{372}  & 90.5          \\
\rowcolor[HTML]{EFEFEF} 
                            &                                                           &                &               &               \\
                            & \multicolumn{1}{l|}{Independent Bayes}                    & 182            & 98.0          & \textbf{94.6} \\
                            & \multicolumn{1}{l|}{Independent Bayes MinT}               & 180            & 97.6          & 93.6          \\
                            & \multicolumn{1}{l|}{Hierarchical Bayes}                   & 181            & 97.4          & 94.4          \\
                            & \multicolumn{1}{l|}{Hierarchical Bayes MinT}              & \textbf{179}   & \textbf{96.4} & 89.7          \\
                            & \multicolumn{1}{l|}{Hierarchical Bayes softer coherency}  & 181            & 97.4          & 94.2          \\
                            & \multicolumn{1}{l|}{Hierarchical Bayes tighter coherency} & 181            & 97.5          & 94.0          \\
                            & \multicolumn{1}{l|}{Hierarchical Bayes total-weighted}    & 181            & 97.2          & 94.2          \\
\multirow{-8}{*}{States}    & \multicolumn{1}{l|}{Hierarchical Bayes purposes-weighted} & 180            & 96.7          & 94.3          \\
\rowcolor[HTML]{EFEFEF} 
                            &                                                           &                &               &               \\
                            & \multicolumn{1}{l|}{Independent Bayes}                    & 79.8           & 41.9          & 95.2          \\
                            & \multicolumn{1}{l|}{Independent Bayes MinT}               & 79.8           & 42.2          & 95.3          \\
                            & \multicolumn{1}{l|}{Hierarchical Bayes}                   & 79.1           & 41.5          & \textbf{95.1} \\
                            & \multicolumn{1}{l|}{Hierarchical Bayes MinT}              & 79.0           & \textbf{41.1} & 93.0          \\
                            & \multicolumn{1}{l|}{Hierarchical Bayes softer coherency}  & 79.0           & 41.5          & \textbf{95.1} \\
                            & \multicolumn{1}{l|}{Hierarchical Bayes tighter coherency} & 79.0           & 41.4          & \textbf{95.1} \\
                            & \multicolumn{1}{l|}{Hierarchical Bayes total-weighted}    & 79.0           & 41.4          & \textbf{95.1} \\
\multirow{-8}{*}{Zones}     & \multicolumn{1}{l|}{Hierarchical Bayes purposes-weighted} & \textbf{78.9}  & 41.4          & \textbf{95.1} \\
\rowcolor[HTML]{EFEFEF} 
                            &                                                           &                &               &               \\
                            & \multicolumn{1}{l|}{Independent Bayes}                    & 41.1           & 21.2          & \textbf{95.6} \\
                            & \multicolumn{1}{l|}{Independent Bayes MinT}               & 41.3           & 21.9          & 96.8          \\
                            & \multicolumn{1}{l|}{Hierarchical Bayes}                   & 40.8           & 21.0          & \textbf{95.6} \\
                            & \multicolumn{1}{l|}{Hierarchical Bayes MinT}              & 40.9           & 21.1          & 95.7          \\
                            & \multicolumn{1}{l|}{Hierarchical Bayes softer coherency}  & \textbf{40.7}  & 21.0          & \textbf{95.6} \\
                            & \multicolumn{1}{l|}{Hierarchical Bayes tighter coherency} & \textbf{40.7}  & \textbf{20.9} & \textbf{95.6} \\
                            & \multicolumn{1}{l|}{Hierarchical Bayes total-weighted}    & \textbf{40.7}  & 21.0          & \textbf{95.6} \\
\multirow{-8}{*}{Regions}   & \multicolumn{1}{l|}{Hierarchical Bayes purposes-weighted} & \textbf{40.7}  & 21.0          & \textbf{95.6}
\end{tabular}
\end{center}
\end{table}

\begin{figure}
    \centering
    \begin{subfigure}[b]{0.48\textwidth}
        \centering
        \includegraphics[width=\textwidth]{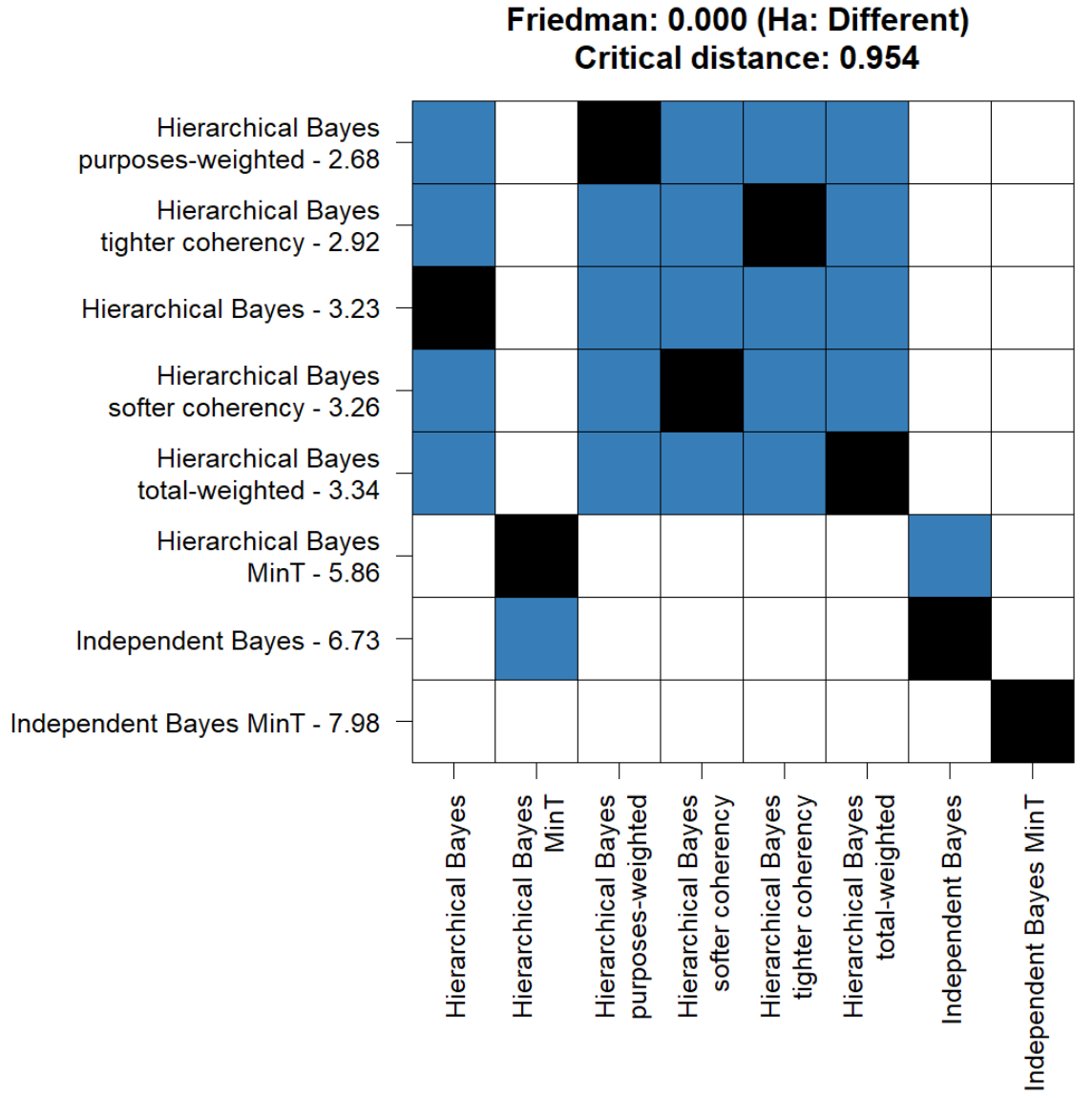}
        \caption{RMSSE}
        \label{fig:RMSSE_nemenyi}
    \end{subfigure}
    \hfill 
    \begin{subfigure}[b]{0.48\textwidth}
        \centering
        \includegraphics[width=\textwidth]{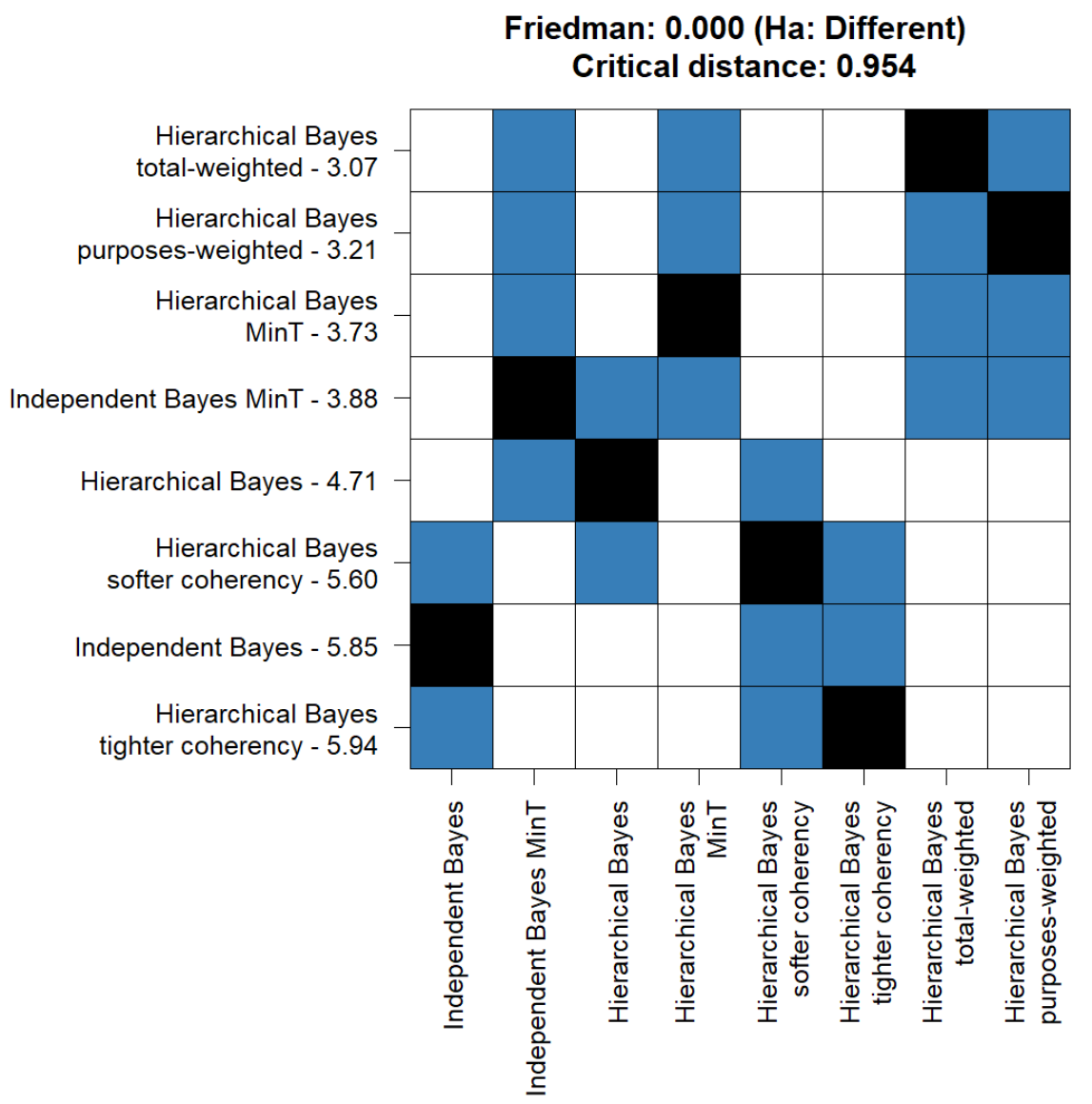}
        \caption{ES}
        \label{fig:CRPS_nemenyi}
    \end{subfigure}
    
    \caption{Nemenyi matrices for the overall accuracy metrics, averaged over all horizons.}
    \label{fig:nemenyi}
\end{figure}

Table \ref{tab:overall_results} shows results for RMSSE and ES for the full hierarchy and RMSE and CRPS averaged over all horizons at each level, and coverage of 95\% prediction intervals for all levels. Relative performance was similar across all horizons, so we report the overall accuracy only. Results for selected individual horizons can be found in the supplementary material Appendix \ref{appendix:tables_tourism}. We see that in the overall results, all methods perform relatively similarly, which is to be expected, as all methods use the same base ETS model structure. However, we do immediately note the efficacy of the Bayesian hierarchical approach, as the two independent Bayesian approaches have the worst overall RMSSE and worse ES than the hierarchical counterparts. The Nemenyi post-hoc test in Figure \ref{fig:nemenyi} verifies this, where the independent Bayes MinT method is ranked worst, demonstrating that the Bayesian hierarchical approach alone is more effective than MinT reconciliation at sharing information between and across hierarchical levels and can improve predictive performance.

The MinT approaches do slightly better with respect to the ES Nemenyi results compared to those for RMSSE, in part due to MinT producing very sharp (under-dispersed) forecasts, but there is no evidence that they are the best performers. 

\begin{figure}[h!]
    \centering
    \includegraphics[width=\linewidth]{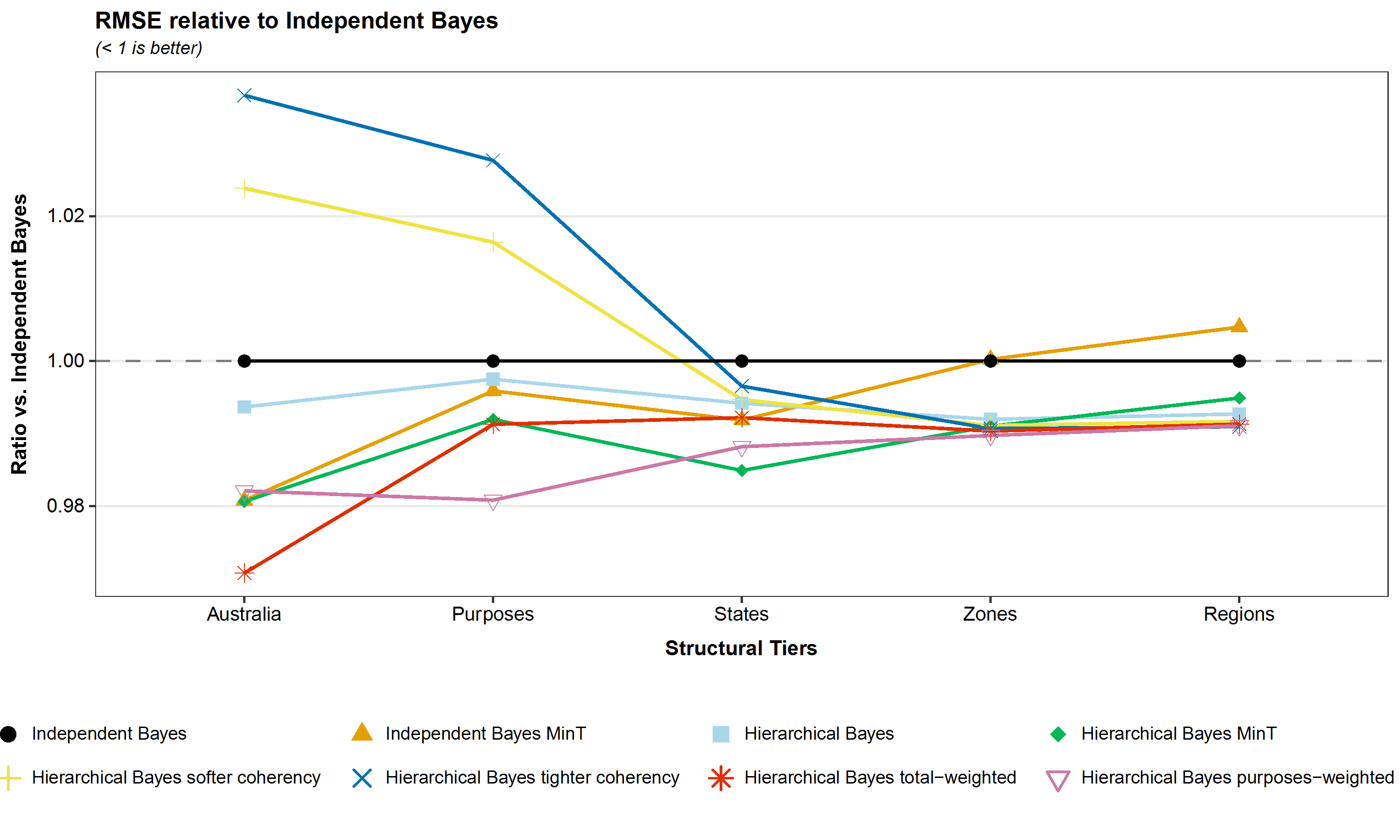}
    \caption{Average RMSE over all horizons at each hierarchical level, split up by method.}
    \label{fig:RMSE_line}
\end{figure}

\begin{figure}[h!]
    \centering
    \includegraphics[width=\linewidth]{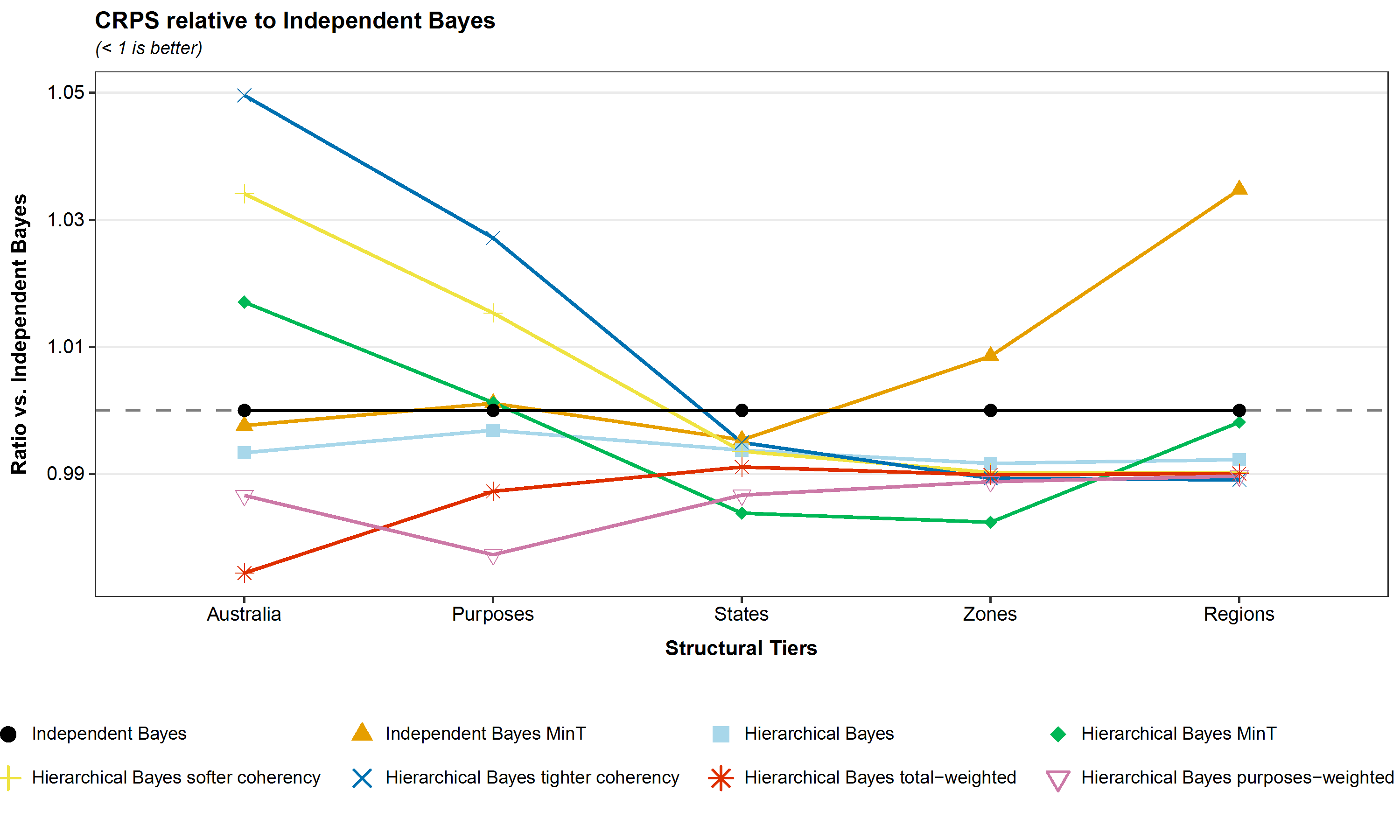}
    \caption{Average CRPS over all horizons at each hierarchical level, split up by method.}
    \label{fig:CRPS_line}
\end{figure}

Figures \ref{fig:RMSE_line} and \ref{fig:CRPS_line} depict the performance of each method relative to the baseline (independent Bayes) at each hierarchical level. The level-weighted approaches are the clear best performers at their respective levels. Furthermore, despite focusing the model on a specific level, the two level-weighted approaches have largely improved performance at other levels, with the Nemenyi results in Figure \ref{fig:nemenyi} demonstrating that the two weighted approaches are the best overall performers.

We also note that the models with our Bayesian coherence penalty alone have better performance compared to baseline and MinT methods at lower hierarchical levels, and slightly poorer performance at upper levels. This behaviour is likely specific to the dataset and model used (as an example, in the paper of \citet{Wickramasuriya2019} with a frequentist ETS model, MinT performed worse at the top level than the unreconciled model). However, this comparatively poor performance at the upper level demonstrates even more clearly the value of our proposed level-focusing approach, obtaining the best overall results at the focused levels with the same coherency penalty applied. 

\begin{table}[h]
\begin{center}
\caption{MAI and JSD between each pair of levels and overall from regions to Australia. MinT methods are excluded from the table as they have no incoherence by construction.} \label{tab:coherency}
\footnotesize
\begin{adjustbox}{max width=\textwidth}
\begin{tabular}{lllllll}
Metric                   & \multicolumn{1}{l|}{Method}                               & \begin{tabular}[c]{@{}l@{}}Australia:\\ Regions\end{tabular} & \begin{tabular}[c]{@{}l@{}}Australia:\\ Purposes\end{tabular} & \begin{tabular}[c]{@{}l@{}}Purposes:\\ States\end{tabular} & \begin{tabular}[c]{@{}l@{}}States:\\ Zones\end{tabular} & \begin{tabular}[c]{@{}l@{}}Zones:\\ Regions\end{tabular} \\ \hline
                         & \multicolumn{1}{l|}{Independent Bayes}                    & 321                                                          & 148                                                           & 197                                                        & 135                                                     & 107                                                      \\
                         & \multicolumn{1}{l|}{Hierarchical Bayes}                   & 286                                                          & 113                                                           & 153                                                        & 119                                                     & 86                                                       \\
                         & \multicolumn{1}{l|}{Hierarchical Bayes softer coherency}  & 119                                                          & 29                                                            & 47                                                         & 63                                                      & 53                                                       \\
                         & \multicolumn{1}{l|}{Hierarchical Bayes tighter coherency} & \textbf{68}                                                  & \textbf{20}                                                   & \textbf{29}                                                & \textbf{36}                                             & \textbf{35}                                              \\
                         & \multicolumn{1}{l|}{Hierarchical Bayes total-weighted}    & 235                                                          & 61                                                            & 77                                                         & 81                                                      & 53                                                       \\      
\end{tabular}
\end{adjustbox}
\end{center}
\end{table}

When assessing coherence in Table \ref{tab:coherency}, we see that when we penalise incoherence, the mean incoherence drops in line with the strength of the penalisation, as predicted. We also see that the hierarchical Bayes approach has a lower mean incoherence compared to the independent Bayes approach, even before penalisation. Mean incoherence is slightly worse despite the penalty in the level-weighted cases, but remains controlled while we have achieved the overall and level-specific gains in accuracy highlighted above. While the coherency penalty improves coherence, the remaining incoherence is reflective of model misspecification, with the use of a Bayesian update that balances accuracy with coherence demonstrating that there is no significantly more coherent solution under the given model specification, as the model did not converge with tighter coherence penalties. This is an indication that our ETS model is misspecified, and a more well-specified model would likely yield both more accurate predictions and more coherent forecasts. In this way, we can see such a penalty as providing a diagnostic tool for model misspecification in hierarchical systems.

We lastly assess the probabilistic calibration of the competing methods in Table \ref{tab:overall_results} by comparing the observed coverage to our 95\% prediction intervals. Here we see that while MinT approaches have no incoherence by construction, they do not have the best calibration at any level and are often the worst. In particular, the upper levels undercover significantly after MinT reconciliation. We note that our level-weighted approaches also undercover slightly at the upper levels -- this is a consequence of the approach, which artificially increases the amount of data, thus making the model overconfident in its predictions. However, this undercoverage remains within 5\% of the target calibration, with overall CRPS scores better than competing methods at these levels. Once again, given a better-specified model, we would expect better calibration overall.

\section{Discussion} \label{discussion}

In this paper, we developed a fully Bayesian hierarchical forecasting framework that shares information between and across levels and ensures soft coherence within model training, with the ability to focus the model on key levels of interest. In this section, we discuss the advantages and disadvantages of our approach and situations when it is applicable. 

In this paper, we first demonstrated that a Bayesian hierarchical approach improves coherence even without reconciliation in both the simulation study and the real data example.  Secondly, in the empirical data example, we see clearly superior performance of the hierarchical Bayes forecasts over the independent Bayesian forecasts, both with and without MinT reconciliation applied. These indicate that a Bayesian hierarchical approach shares information between series more effectively than post-hoc MinT reconciliation alone.

However, Bayesian hierarchical approaches help to estimate model parameters when they are: i) challenging to identify from the given data; ii) when it is known that series share characteristics.  The first case may result when there is unclear, little or missing data, improving parameter estimates by drawing uncertain estimates toward the population hyperprior, as illustrated in the simulation study example where we varied training data length. The Australian tourism forecasting example demonstrates a more realistic scenario where the lower levels likely have less reliable data. In this case, the hierarchical structure helps identify the parameters of each series and improves performance, particularly at the lowest two levels, where all the Bayesian hierarchical approaches outperform the two independent approaches. To address the second point, a Bayesian hierarchical approach is not appropriate if series do not share characteristics or if different parametric models are used for different series. In this case, our coherency penalty and level weightings still apply and will act to produce softly coherent forecast that focus on performance at a particular level or series. However, we do not recommend using the Bayesian hierarchical structure on parameters to directly share information between levels, so we would not necessarily expect to see the same magnitude of benefit from focusing the model on levels with better data quality. 

Our approach to softly penalise incoherence provides a flexible balance between hierarchical coherence and model fit. Empirically, we have seen that we can obtain good mean coherence regardless of model specification, although incoherence between levels further apart  increases, and we were not able to obtain parameters that yielded perfect coherence. This is in contrast to MinT reconciliation, which does not account for the level of model misspecification and results in fully coherent, but poorly calibrated forecasts. Our approach is superior in terms of maintaining probabilistic calibration at upper levels while significantly reducing mean incoherence. If strict coherence is required, an approach that guarantees probabilistic coherence would be more appropriate, but such forecasts may be poorly calibrated, resulting in poor decisions. In scenarios where a small degree of incoherence is permissible, our approach provides good coherence balanced against model fit, and may indicate when model misspecification is limiting coherence by lack of convergence under a strict incoherence penalty, allowing the forecaster to re-evaluate model choice. 

The focusing technique we develop increases the weight of a particular level or series relative to the other series in the hierarchy. This increased weight means that predictions at this level will be less affected by reconciliation and will also have a higher weight within the Bayesian hierarchy on parameters, drawing the parameters of related series closer to the parameters of the focused level. In the Australian tourism example, we see that in addition to this level being less affected by reconciliation, this weighting can provide considerable benefit when the model or data at an upper level is good. However, the simulation study illustrates the opposite case: when the model at the upper level is more misspecified, focusing on this level worsens overall predictions across the hierarchy. Consequently, although a particular level might be more relevant to decision-making and the base forecast more accurate than the reconciled forecast, focusing may in fact reduce performance if the data or model are poor. This decision can be made based on model selection criteria relative to a baseline model or predictive performance to avoid focusing incorrectly. 

Although the focusing approach potentially improves the results for the focused level and related hierarchical levels, it will come at the cost of some posterior calibration of the parameters (the uncertainty associated with the parameters in the Bayesian framework), and, in small sample sizes, predictive distribution calibration. In the literature for generalised Bayesian updating, there are a number of methods for maintaining calibration of posteriors \citep{Bissiri2016}, but applying these would largely counteract our goal of focusing the model in order to optimise the original log-likelihood update. We refer to recent work in predictive generalised Bayesian approaches \citep{McLatchie2025}, which show that in asymptotically large sample sizes, predictive inference is not significantly affected by the posterior parameter miscalibration. In small or moderate sample sizes, we accept that some miscalibration will likely occur, but this disadvantage is offset by the improved focus on that level within the full model. In this paper, we have demonstrated empirically that positive results can be obtained by weighting, but further work is needed to determine precise strategies for balancing the effect of focusing against calibration.

Lastly, a limitation of our framework is computational scalability, as we fit our models using MCMC sampling. While MCMC methods are considered the gold standard for accurately sampling Bayesian posteriors, they are computationally expensive. Having now demonstrated that the key aspects of our methodology are effective, future work is needed to enhance the scalability of this framework. One option is to use conjugate priors where available, allowing us to obtain the posterior in closed form. However, this approach does not immediately generalise to incorporate our coherency penalty or level weighting, and many statistical models do not have conjugate priors. A second option is to explore approximate Bayesian techniques, such as variational inference \citep{Blei2017}, where the posterior is approximated using a simpler class of models. Alternatively, for small hierarchies, or moderately sized hierarchies with limited data availability, MCMC sampling is likely to be feasible.

\color{black}

\section{Conclusion} \label{conclusion}
We have introduced the use of Bayesian hierarchies to perform hierarchical forecasting, demonstrating how they can improve predictive performance when series are known to be related. In addition to this, we have proposed and implemented a global coherency framework that allows us to penalise mean incoherence within model training, without the need to estimate covariance matrices for multi-step forecast horizons. We also combine this with an approach to weight levels or series that have better data or are of greater relevance to decision-making, focusing the reconciled hierarchy on the predictions at that level. 

We have demonstrated the benefit of our approach in forecasting Australian domestic tourism, outperforming MinT reconciliation in point and probabilistic accuracy metrics overall and at the focused levels. Our incoherence penalty finds a log-likelihood-optimal solution balancing accuracy against coherence in misspecified models. This is an improvement over reconciliation methods that force coherence at the expense of model specification and with consequences for forecast calibration. 

Our framework is highly flexible, and future work could include extending it to cross-temporal and multivariate hierarchies. We also expect that the Bayesian hierarchical structure could be even better leveraged by combining with clustering methods, as in the non-Bayesian approach of \citep{Zhang2025}, as this would allow the data to inform which series would most benefit from information pooling. When using weighting to reward accuracy for a given series or level, our approach could be combined with that of \citet{Hollyman2022}, whose method identifies which series in the hierarchy contribute most to improving accuracy following reconciliation. Additionally, while in this paper we used simple ARMA and ETS models, our methodology can be applied to fully Bayesian time series models, such as those described in \citet{Prado2021}, which may further improve forecasting performance. Using conjugate priors, variational inference or approximate Bayesian computation techniques would also be a promising avenue to explore to reduce the computational cost of training Bayesian forecasts. It would finally be of interest to investigate combining our level-focusing approach with methods that generate fully coherent forecasts.

\section{Code}
\url{https://github.com/ArwenNugteren/hierarchical_bayes}

\section*{Acknowledgements}
The authors thank two anonymous reviewers and an associate editor who provided feedback on this manuscript. 

\section*{Funding}
Arwen Nugteren is supported by an Australian Government Research Training Program Scholarship:  \url{https://doi.org/10.82133/C42F-K220}. Christopher Drovandi is supported by the Australian Research Council. 

\section*{Author contributions}
Arwen Nugteren: Conceptualisation; Formal analysis; Methodology; Investigation; Writing - original draft; Writing - review \& editing. 
Mahdi Abolghasemi: Conceptualisation; Supervision; Writing - review \& editing.
Kerrie Mengersen: Conceptualisation; Supervision; Writing - review \& editing.
Christopher Drovandi: Supervision; Writing - review \& editing.



\appendix

\section{Simulation study additional details} \label{appendix:simulation}

\color{black}
We employ a similar simulation set-up to that used in \citet{Wickramasuriya2019} and \citet{Corani2021}, with a three-level hierarchy separated into 4 bottom-level series, 2 middle-level aggregate series and one total series, as shown in Figure \ref{fig:hierarchy_diagram}. 

We first generate four series $z_{t,i}$ for $i \in \{AA, AB, BA, BB\}$ from an ARMA(2,2) model with parameters $\phi_1, \phi_2, \theta_1$ and $\theta_2$,  
\begin{equation}
    z_{i,t} = \phi_1 z_{i,t-1} + \phi_2 z_{i,t-2} + \theta_1 \varepsilon_{i,t-1} + \theta_2 \varepsilon_{i,t-2} + \varepsilon_{i,t},
\end{equation}

where the noise process is multivariate Gaussian, $(\bm{\varepsilon}_{t,AA}, \bm{\varepsilon}_{t,AB}, \bm{\varepsilon}_{t,BA}, \bm{\varepsilon}_{t,BB}) \sim N(\bm{0}, \Sigma)$, with the covariance matrix 
\begin{equation*}
    \Sigma = \begin{pmatrix}
        5 & 3 & 2 & 1 \\ 
        3 & 4 & 2 & 1 \\
        2 & 2 & 5 & 3 \\
        1 & 1 & 3 & 4
    \end{pmatrix},
\end{equation*}
which specifies a stronger correlation between the bottom-level series that share the same parent node than those that do not. As noted in \citet{Wickramasuriya2019}, it is typical that upper-level series have less noise, as aggregation smooths out random fluctuations. To model this behaviour in this small and synthetic hierarchy, we adopt the same approach as in their paper, adding and subtracting two noise terms $\nu_t$ and $\omega_t$ \color{black}to each ARMA(2,2) series ($z_{AA,t}, z_{AB,t}, z_{BA,t}, z_{BB,t}$) as follows:
\begin{align*}
    y_{AA,t} & = z_{AA,t} - \nu_t - 0.5\omega_t, \\
    y_{AB,t} & = z_{AB,t} + \nu_t - 0.5\omega_t, \\
    y_{BA,t} & = z_{BA,t} - \nu_t + 0.5\omega_t, \\
    y_{BB,t} & = z_{BB,t} + \nu_t + 0.5\omega_t,
\end{align*}
such that the $\nu_t$ terms cancel at the middle level (series $A$ and $B$) and the $\omega_t$ terms cancel at the top level. We specify independent error process $\nu_t \sim N(0, 10)$ and $\omega_t \sim N(0,6)$ as is done in \citet{Wickramasuriya2019}.

We sample the bottom-level parameters from a hierarchical structure where the ARMA parameters of series $AA$ and $AB$ are drawn from a Gaussian hierarchical prior at level $A$ with standard deviation 0.1, and the parameters from series $BA$ and $BB$ are drawn from a Gaussian hierarchical prior at level $B$ with standard deviation 0.1. The ARMA parameters of the hyperpriors corresponding to series $A$ and $B$ are drawn from a second level of the Gaussian hyperprior, representing the upper level, also given a standard deviation of 0.1. The hyperparameters of the top-level hyperprior are drawn uniformly from $\phi_2, \theta_2 \in [0.5,0.7]$, $\phi_1 \in [\phi_2 - 0.9, 0.9-\phi_2]$ and $\theta_1 \in [-(0.9+\theta_2)/3.2, (0.9+\theta_2)/3.2]$. This choice of parameter space follows that of \citet{Wickramasuriya2019} and ensures stationarity and invertibility. 

We then simulate a total of 755 observations, discarding the first 200 as burn-in for the ARMA process, leaving 555 data points for our experiments, where we have a training window size of up to 50 observations, produce forecasts for horizons $h=1, \ldots, 5$, and perform a rolling window evaluation over 500 windows.

\color{black}

\subsection{Well-specified model}
We demonstrate the efficacy of our modelling approach in the case where the bottom level of the model is both well-specified and misspecified. For the well-specified model, we choose an ARMA(2,2) model with the bottom-level series modelled with a joint covariance matrix. The upper levels, $A$, $B$ and the total are also modelled using ARMA(2,2) processes with additive Gaussian noise. 

We explore two methods of learning the ARMA parameters -- independent estimation and Bayesian hierarchical estimation. For the independent Bayes (baseline) models, all ARMA parameters are given $U(-1,1)$ priors, covering the entire parameter space. Bayesian fitting of ARMA models is well-established as an effective time series forecasting technique \citep{Prado2021}.

For the Bayesian hierarchical models, we structure the parameter hierarchy to align with the aggregate hierarchy. This means that the parameters for series $AA$, $AB$ and $A$ are all drawn from the same Gaussian hyperprior with standard deviation 0.1, with the hyperprior mean drawn from a second Gaussian hyperprior level shared with the hyperprior mean for both the $B$ series and the total series, also given standard deviations of 0.1. This top-level shared hyperprior has hyperparameters given $U(-1,1)$ prior distributions. If we let $\bm{\theta}_i$ be a vector containing all the ARMA parameters for series $i$, with hyperprior means for level $j$ given by $\bm{\mu}_j$, then we have the following priors and hyperpriors: 
\begin{align}
    \bm{\theta}_{AA}, \bm{\theta}_{AB}, \bm{\theta}_{A} & \sim N(\bm{\mu}_A, 0.1^2 I_{4\times 4}) \\
    \bm{\theta}_{BA}, \bm{\theta}_{BB}, \bm{\theta}_{B} & \sim N(\bm{\mu}_B, 0.1^2 I_{4\times 4}) \\
    \bm{\mu}_{A}, \bm{\mu}_{B}, \bm{\theta}_\mathrm{top} & \sim N(\bm{\mu}_\mathrm{total}, 0.1^2 I_{4\times 4}) \\
    \bm{\mu}_\mathrm{total} & \sim U(-1,1)^4.
\end{align}

\color{black}
Variance parameters are given $\text{half-Cauchy}(0,1)$ priors, and the bottom-level covariance is given a Cholesky LKJ correlation prior with shape parameter 2, as recommended in the Stan manual for multivariate Gaussian models \citep{stan_users_guide_2026}. The prior hyperparameters were chosen to yield weakly informative priors. Prior predictive checks were performed for both the independent and Bayesian hierarchical models and are reported for a representative series at each level in the supplementary Figures \ref{fig:prior_predictive_independent} and \ref{fig:prior_predictive_hierarchical}. 

The likelihood for series $i$ is then given by
\begin{equation}
    y_{i,t} = \phi_1 y_{i,t-1} + \phi_2 y_{i,t-2} + \theta_1 \varepsilon_{i,t-1} + \theta_2 \varepsilon_{i,t-2} + \varepsilon_{i,t}, 
\end{equation}
where the error process $\varepsilon_{i,t}$ is drawn from either the 4-dimensional Gaussian distribution at the bottom level or from a univariate Gaussian at the aggregate levels. 

\color{black}
\subsection{Misspecified model}
We also show the behaviour of the method in the case where the model is known to be misspecified. We choose an additive exponential smoothing (ETS) model with level and trend \citep{Holt1957}, and use independent Laplace distributions for the errors. This choice means that the model will not account for the known temporal correlation in the error process and uses an incorrect and independent distribution for the probabilistic model. Laplace distributions also do not satisfy the property of aggregating to another Laplace distribution, introducing a further source of misspecification. Using Bayesian fitting for ETS models with non-Gaussian error distributions has been established as effective in the literature \citep{Smyl2025}. 

This gives the general form of the likelihood for a given series as, 
\begin{align*}
    y_{t+h \vert h} & = \ell_t + hb_t + \varepsilon_t, \\
    \ell_t & = \ell_{t-1} + b_{t-1} + \alpha \varepsilon_t, \\
    b_t & = b_{t-1} + \beta \varepsilon_t, \\
    \varepsilon_t & \sim Laplace(0,\sigma).
\end{align*}
Here, we use $\sigma$ as the scale parameter of the Laplace distribution, although we note that this will be different from the standard deviation of the Laplace distribution. We then put priors on $\alpha$ and $\beta$. In the independent Bayesian case, we place $U(0,1)$ priors on both $\alpha$ and $\beta$. We also need to specify priors for the level and trend, which we choose to be $\ell_0 \sim N(0, 10^2)$ and $b_0 \sim N(0, 5^2)$. We place $\text{half-Cauchy}(0,1)$ priors on the variances at each level. 

In the hierarchical Bayesian case, we follow a similar structure to the well-specified case. Letting $\bm{\theta}_i$ be a vector containing the parameters $\alpha$ and $\beta$ for series $i$, we have: 

\begin{align*}
    \bm{\theta}_{AA}, \bm{\theta}_{AB}, \bm{\theta}_{A} & \sim N(\bm{\mu}_A, 0.1^2 I_{2\times 2}) \\
    \bm{\theta}_{BA}, \bm{\theta}_{BB}, \bm{\theta}_{B} & \sim N(\bm{\mu}_B, 0.1^2 I_{2\times 2}) \\
    \bm{\mu}_{A}, \bm{\mu}_{B}, \bm{\theta}_\mathrm{top} & \sim N(\bm{\mu}_\mathrm{total}, 0.1^2 I_{2\times 2}) \\
    \bm{\mu}_\mathrm{total} & \sim U(0,1)^2.
\end{align*}

We then give the same priors on initial states, $\ell_0 \sim N(0, 10^2)$ and $b_0 \sim N(0, 5^2)$, and place $\text{half-Cauchy}(0,1)$ priors on the variances at each level.

\subsection{Competing methods}
We compare with the MinT (sample) method of \citet{Panagiotelis2023} for probabilistic forecast reconciliation on both the independent and hierarchical Bayes trials. We use the shrinkage estimator of the covariance matrix and use the approximation $\widehat{W}_h = k_h \widehat{W_1}$ with $k_h=1$, where the residuals are calculated using the posterior mean fitted values for the 1-step-ahead forecasts.

\subsection{Computational details}
All models were fit in Stan, running 10 chains with 2000 samples each, discarding the first 1000 as warm-up, for a total of 10,000 posterior samples. Individual window model training and sampling required less than 2 minutes on a standard laptop computer, with coherency-penalised trials taking slightly longer than their non-penalised counterparts. To run the full 500 rolling window evaluation, a high-performance computing system was used with trials batched and run in parallel.

\subsection{Results}
In the well-specified case, the simulation study represents a case where the error process is a simple Gaussian distribution, so it is simple to estimate the parameters, and an independent Bayesian approach is sufficient to capture the model structure. The use of a parameter hierarchy shrinks estimates towards the hierarchical mean -- further from the true posterior mean in such cases. This is demonstrated in Table \ref{tab:appendix_training_window} for the varying training window length, where knowledge of the true data-generating process and simple errors means there is little to no advantage of using the Bayesian hierarchical structure.

\begin{table}[h!]
\caption{Results for overall accuracy comparing the well-specified and misspecified trials under varying training window length. } \label{tab:appendix_training_window}
\begin{adjustbox}{max width=\textwidth}
\begin{tabular}{lllllllll}
Training data size   &                                             & Well-specified           &                          &                          &                          & Misspecified             &                          &                          \\
                     & \multicolumn{1}{l|}{Method}                 & RMSSE                    & ES                       & 95\% Coverage            &                          & RMSSE                    & ES                       & 95\% Coverage            \\ \cline{2-9} 
                     & \cellcolor[HTML]{EFEFEF}                    & \cellcolor[HTML]{EFEFEF} & \cellcolor[HTML]{EFEFEF} & \cellcolor[HTML]{EFEFEF} & \cellcolor[HTML]{EFEFEF} & \cellcolor[HTML]{EFEFEF} & \cellcolor[HTML]{EFEFEF} & \cellcolor[HTML]{EFEFEF} \\
                     & \multicolumn{1}{l|}{Independent Bayes}      & \textbf{0.59}            & \textbf{11.3}            & \textbf{93.0}            &                          & 0.82                     & 16.8                     & 98.4                     \\
                     & \multicolumn{1}{l|}{Hierarchical Bayes}     & \textbf{0.59}            & \textbf{11.3}            & 91.5                     &                          & \textbf{0.79}            & \textbf{16.0}            & \textbf{97.3}            \\
\multirow{-3}{*}{10} & \multicolumn{1}{l|}{Percentage improvement} & -0.99\%                  & 0.24\%                   & -1.53\%                  &                          & 3.82\%                   & 4.75\%                   & 1.07\%                   \\
                     &                                             &                          &                          &                          &                          &                          &                          &                          \\
                     & \multicolumn{1}{l|}{Independent Bayes}      & \textbf{0.58}            & \textbf{11.0}            & \textbf{94.2}            &                          & 0.75                     & 15.0                     & 98.5                     \\
                     & \multicolumn{1}{l|}{Hierarchical Bayes}     & 0.59                     & 11.1                     & 93.4                     &                          & \textbf{0.74}            & \textbf{14.7}            & \textbf{97.6}            \\
\multirow{-3}{*}{20} & \multicolumn{1}{l|}{Percentage improvement} & -1.74\%                  & -0.94\%                  & -0.85\%                  &                          & 1.13\%                   & 1.55\%                   & 0.84\%                   \\
                     &                                             &                          &                          &                          &                          &                          &                          &                          \\
                     & \multicolumn{1}{l|}{Independent Bayes}      & \textbf{0.57}            & \textbf{10.9}            & \textbf{95.0}            &                          & \textbf{0.70}            & 14.0                     & 97.7                     \\
                     & \multicolumn{1}{l|}{Hierarchical Bayes}     & 0.58                     & 11.0                     & 94.9                     &                          & \textbf{0.70}            & \textbf{13.9}            & \textbf{97.5}            \\
\multirow{-3}{*}{50} & \multicolumn{1}{l|}{Percentage improvement} & -1.38\%                  & -0.72\%                  & -0.12\%                  &                          & 0.24\%                   & 0.45\%                   & 0.37\%                  
\end{tabular}
\end{adjustbox}
\end{table}

Table \ref{tab:appendix_simulation_accuracy} also shows that for the well-specified model, the independent Bayesian models are largely the best performers in terms of RMSE and CRPS. This is the result of correct model specification, where the independent Bayesian update is the optimal method for updating information to a posterior \citep{Zellner1988}. As the base model is correctly specified, adding incoherence penalties and level weightings slightly misspecifies the model by shrinking parameter estimates towards the group mean, hence yields comparatively worse results. However, we do see that our Bayesian coherency approach improves over MinT approaches in terms of coverage. The focused methods improve CRPS scores over the Bayes coherent approaches with the same coherency penalty at their respective levels.

\begin{table}[h!]
\caption{Overall accuracy results at all levels for both the well-specified and misspecified models}.  \label{tab:appendix_simulation_accuracy}
\footnotesize
\begin{adjustbox}{max width=\textwidth}
\begin{tabular}{lllllllll}
                          &                                                           & \multicolumn{3}{l}{\textbf{Well-specified}}   &  & \multicolumn{3}{l}{\textbf{Misspecified}}      \\
Level                     & \multicolumn{1}{l|}{Method}                               & RMSSE         & ES            & 95\% Coverage &  & RMSSE          & ES            & 95\% Coverage \\ \hline
\rowcolor[HTML]{EFEFEF} 
                          &                                                           &               &               &               &  &                &               &               \\
                          & \multicolumn{1}{l|}{Independent Bayes}                    & \textbf{0.57} & 10.9          & \textbf{95.0} &  & \textbf{0.70}  & 14.0          & 97.7          \\
                          & \multicolumn{1}{l|}{Independent Bayes MinT}               & \textbf{0.57} & \textbf{10.7} & 87.5          &  & \textbf{0.70}  & \textbf{13.7} & 91.0          \\
                          & \multicolumn{1}{l|}{Hierarchical Bayes}                   & 0.58          & 11.0          & 94.9          &  & \textbf{0.70}  & 13.9          & 97.5          \\
                          & \multicolumn{1}{l|}{Hierarchical Bayes MinT}              & 0.58          & 10.9          & 87.0          &  & \textbf{0.70}  & \textbf{13.7} & 90.3           \\
                          & \multicolumn{1}{l|}{Hierarchical Bayes softer coherency}  & 0.59          & 11.1          & 94.9          &  & \textbf{0.70}  & 13.9          & 97.1          \\
                          & \multicolumn{1}{l|}{Hierarchical Bayes tighter coherency} & 0.59          & 11.1          & 94.8          &  & \textbf{0.70}  & 13.9          & 97.1          \\
\multirow{-7}{*}{Overall} & \multicolumn{1}{l|}{Hierarchical Bayes total-weighted}    & 0.60          & 11.2          & 95.1          &  & \textbf{0.70}  & 14.0          & \textbf{96.9} \\
                          & \multicolumn{1}{l|}{Hierarchical Bayes bottom-weighted}   & 0.59          & 11.1          & 94.7          &  & \textbf{0.70}  & 13.8          & \textbf{96.9} \\
                          &                                                           &               &               &               &  &                &               &               \\
Level                     & Method                                                    & RMSE          & CRPS          & 95\% Coverage &  & RMSE           & CRPS          & 95\% Coverage \\ \cline{3-5} \cline{7-9} 
\rowcolor[HTML]{EFEFEF} 
                          &                                                           &               &               &               &  &                &               &               \\
                          & \multicolumn{1}{l|}{Independent Bayes}                    & 9.26          & \textbf{5.26} & \textbf{94.0} &  & \textbf{12.50} & \textbf{7.15} & 98.0          \\
                          & \multicolumn{1}{l|}{Independent Bayes MinT}               & \textbf{9.27} & 5.53          & 75.4          &  & 12.52          & 7.40          & 80.5          \\
                          & \multicolumn{1}{l|}{Hierarchical Bayes}                   & 9.34          & 5.31          & 92.7          &  & 12.61           & 7.21            & 97.2            \\
                          & \multicolumn{1}{l|}{Hierarchical Bayes MinT}              & 9.35          & 5.62          & 72.6          &  & 12.61           & 7.49            & 79.4           \\
                          & \multicolumn{1}{l|}{Hierarchical Bayes softer coherency}  & 9.44          & 5.38          & 92.0          &  & 12.84            & 7.31          & 96.0            \\
                          & \multicolumn{1}{l|}{Hierarchical Bayes tighter coherency} & 9.44          & 5.39          & 92.0          &  & 12.85            & 7.31          & 95.9            \\
\multirow{-7}{*}{Top}     & \multicolumn{1}{l|}{Hierarchical Bayes total-weighted}    & 9.34          & 5.33          & 91.3          &  & 12.94            & 7.40          & \textbf{95.4} \\
                          & \multicolumn{1}{l|}{Hierarchical Bayes bottom-weighted}   & 9.45          & 5.39          & 91.7          &  & 12.77            & 7.26          & 96.1            \\
\rowcolor[HTML]{EFEFEF} 
                          &                                                           &               &               &               &  &                &               &               \\
                          & \multicolumn{1}{l|}{Independent Bayes}                    & 6.45          & 3.65          & \textbf{94.9} &  & 8.37           & 4.79          & 97.5          \\
                          & \multicolumn{1}{l|}{Independent Bayes MinT}               & \textbf{6.31} & 3.64          & 82.4          &  & \textbf{8.26}  & 4.79          & 87.3          \\
                          & \multicolumn{1}{l|}{Hierarchical Bayes}                   & 6.42          & \textbf{3.62} & 94.2          &  & 8.32             & \textbf{4.77} & 97.4            \\
                          & \multicolumn{1}{l|}{Hierarchical Bayes MinT}              & 6.39          & 3.71          & 81.0          &  & 8.28           & 4.81           & 86.2            \\
                          & \multicolumn{1}{l|}{Hierarchical Bayes softer coherency}  & 6.44          & 3.63          & 94.1          &  & 8.37           & 4.79          & 97.0          \\
                          & \multicolumn{1}{l|}{Hierarchical Bayes tighter coherency} & 6.45          & 3.64          & 94.1          &  & 8.38           & 4.80          & 96.9            \\
\multirow{-7}{*}{Middle}  & \multicolumn{1}{l|}{Hierarchical Bayes total-weighted}    & 6.47          & 3.65          & 94.5          &  & 8.38           & 4.80          & \textbf{96.7} \\
                          & \multicolumn{1}{l|}{Hierarchical Bayes bottom-weighted}   & 6.44          & 3.64          & 94.0          &  & 8.33             & \textbf{4.77} & 97.1          \\
\rowcolor[HTML]{EFEFEF} 
                          &                                                           &               &               &               &  &                &               &               \\
                          & \multicolumn{1}{l|}{Independent Bayes}                    & 4.95          & 2.81          & \textbf{95.4} &  & 5.86           & 3.35          & 97.8          \\
                          & \multicolumn{1}{l|}{Independent Bayes MinT}               & \textbf{4.92} & \textbf{2.80} & 93.0          &  & 5.84           & 3.33          & 95.5          \\
                          & \multicolumn{1}{l|}{Hierarchical Bayes}                   & 5.06          & 2.88          & 95.9          &  & 5.84           & 3.34          & 97.6            \\
                          & \multicolumn{1}{l|}{Hierarchical Bayes MinT}              & 5.06          & 2.88          & 93.7          &  & 5.82           & 3.33          & \textbf{95.1} \\
                          & \multicolumn{1}{l|}{Hierarchical Bayes softer coherency}  & 5.12          & 2.92          & 96.0          &  & 5.81           & \textbf{3.32} & 97.5          \\
                          & \multicolumn{1}{l|}{Hierarchical Bayes tighter coherency} & 5.14          & 2.93          & 95.9          &  & 5.81            & 3.33          & 97.5           \\
\multirow{-7}{*}{Bottom}  & \multicolumn{1}{l|}{Hierarchical Bayes total-weighted}    & 5.22          & 2.99          & 96.4          &  & 5.82           & 3.33          & 97.3           \\
                          & \multicolumn{1}{l|}{Hierarchical Bayes bottom-weighted}   & 5.12          & 2.92          & 95.9          &  & \textbf{5.79}  & \textbf{3.32} & 97.0         
\end{tabular}
\end{adjustbox}
\end{table}

When comparing incoherence penalisation between the model specifications in Table \ref{tab:appendix_simulation_coherence}, the amount of incoherence increases between levels that are further apart in the misspecified case. This is not observed in the well-specified case, where incoherence is expected to asymptotically vanish due to the correct model specification. Overall, these results demonstrate that we can obtain substantial improvements to mean coherence regardless of model specification.

\begin{table}[h!]
\centering
\caption{Coherency performance for the well-specified and misspecified models, excluding MinT reconciliation methods, which have no incoherence by construction.} \label{tab:appendix_simulation_coherence}
\footnotesize
\begin{adjustbox}{max width=\textwidth}
\begin{tabular}{lllllllll}
                      &                                      & \multicolumn{3}{l}{\textbf{Well-specified}}                                                                                                                              &  & \multicolumn{3}{l}{\textbf{Misspecified}}                                                                                                                                \\
Metric                & Method                               & \begin{tabular}[c]{@{}l@{}}Top:\\ Bottom\end{tabular} & \begin{tabular}[c]{@{}l@{}}Top:\\ Middle\end{tabular} & \begin{tabular}[c]{@{}l@{}}Middle:\\ Bottom\end{tabular} &  & \begin{tabular}[c]{@{}l@{}}Top:\\ Bottom\end{tabular} & \begin{tabular}[c]{@{}l@{}}Top:\\ Middle\end{tabular} & \begin{tabular}[c]{@{}l@{}}Middle:\\ Bottom\end{tabular} \\
\rowcolor[HTML]{EFEFEF} 
                      &                                      &                                                       &                                                       &                                                          &  &                                                       &                                                       &                                                          \\
MAI                   & Independent Bayes                    & 2.87                                                  & 1.98                                                  & 2.01                                                     &  & 2.09                                                  & 1.69                                                  & 2.01                                                     \\
                      & Hierarchical Bayes                   & 1.24                                                  & 0.91                                                  & 0.71                                                     &  &   1.74                                                  &  1.28                                                  &   1.48                                                     \\
                      & Hierarchical Bayes softer coherency  & 0.30                                                  & 0.17                                                  & 0.22                                                     &  &  0.21                                                    &  0.16                                                    & 0.19                                                     \\
                      & Hierarchical Bayes tighter coherency & \textbf{0.21}                                         & \textbf{0.13}                                         & \textbf{0.16}                                            &  & \textbf{0.18}                                         & \textbf{0.15}                                         & \textbf{0.16}                                            \\
                      & Hierarchical Bayes total-weighted    & 0.23                                                  & 0.14                                                  & 0.17                                                     &  & 0.22                                                  & 0.17                                                  & 0.18                                            \\
                      & Hierarchical Bayes bottom-weighted   & 0.35                                                  & 0.17                                                  & 0.26                                                     &  & 0.20                                                  & 0.15                                                  & 0.18                                                     \\      
\end{tabular}
\end{adjustbox}
\end{table}

Lastly, we report the results for the well-specified model on the incoherent data in Tables \ref{tab:appendix_incoherent_data} and \ref{tab:appendix_incoherent_data_MAI} for completeness. In this case, we see little difference between the model run on coherent and incoherent data, with slightly better performance on the coherent data. This is likely a result of the model not being strictly ``well-specified'' given the change to the data-generating process.  

\begin{table}[h]
\caption{Results for overall accuracy comparing the well-specified model's performance on the coherent and incoherent data. } \label{tab:appendix_incoherent_data}
\begin{adjustbox}{max width=\textwidth}
\begin{tabular}{llllllll}
                                                          & \multicolumn{3}{l}{Coherent data}             &  & \multicolumn{3}{l}{Incoherent data}           \\
\multicolumn{1}{l|}{Method}                               & RMSSE         & ES            & 95\% Coverage &  & RMSSE         & ES            & 95\% Coverage \\ \hline
\rowcolor[HTML]{EFEFEF} 
                                                          &               &               &               &  &               &               &               \\
\multicolumn{1}{l|}{Independent Bayes}                    & \textbf{0.57} & 10.9          & \textbf{95.0} &  & 0.58          & \textbf{95.0} & 11.0          \\
\multicolumn{1}{l|}{Independent Bayes MinT}               & \textbf{0.57} & \textbf{10.7} & 87.5          &  & \textbf{0.57} & 87.4          & \textbf{10.8} \\
\multicolumn{1}{l|}{Hierarchical Bayes}                   & 0.58          & 11.0          & 94.9          &  & 0.58          & 94.8          & 11.1          \\
\multicolumn{1}{l|}{Hierarchical Bayes MinT}              & 0.58          & 10.9          & 87.0          &  & 0.58          & 86.8          & 11.0          \\
\multicolumn{1}{l|}{Hierarchical Bayes softer coherency}  & 0.59          & 11.1          & 94.9          &  & 0.58          & 93.3          & 11.2          \\
\multicolumn{1}{l|}{Hierarchical Bayes tighter coherency} & 0.59          & 11.1          & 94.8          &  & 0.59          & 93.0          & 11.4         
\end{tabular}
\end{adjustbox}
\end{table}

\begin{table}[h]
\caption{Mean absolute incoherence results for the well-specified model on the incoherent data.  } \label{tab:appendix_incoherent_data_MAI}
\begin{adjustbox}{max width=\textwidth}
\begin{tabular}{llllllll}
                                                          & \multicolumn{3}{l}{Coherent data}             &  & \multicolumn{3}{l}{Incoherent data}           \\
\multicolumn{1}{l|}{Method}                               & Top:Bottom    & Top:Middle    & Middle:Bottom &  & Top:Bottom    & Top:Middle    & Middle:Bottom \\ \hline
\rowcolor[HTML]{EFEFEF} 
Raw data                                                  & 0             & 0             & 0             &  & 1.16          & 1.32          & 1.27          \\
\multicolumn{1}{l|}{Independent Bayes}                    & 2.87          & 1.98          & 2.01          &  & 3.06          & 2.28          & 2.20          \\
\multicolumn{1}{l|}{Hierarchical Bayes}                   & 1.24          & 0.91          & 0.71          &  & 1.46          & 1.25          & 1.05          \\
\multicolumn{1}{l|}{Hierarchical Bayes softer coherency}  & 0.30          & 0.17          & 0.22          &  & 0.51          & 0.53          & 0.51          \\
\multicolumn{1}{l|}{Hierarchical Bayes tighter coherency} & \textbf{0.21} & \textbf{0.13} & \textbf{0.16} &  & \textbf{0.40} & \textbf{0.43} & \textbf{0.41}
\end{tabular}
\end{adjustbox}
\end{table}

\section{Prior predictive checks}
In this section, we include plots of the prior predictive against the data in the first training window. For each model, we pick a representative series from each hierarchical level, as all series have the same priors and hyperpriors, with the exception of a scaling term based on the number of series for the tourism example. All prior predictives were generated from 10,000 samples. 

\begin{figure}
    \centering
    \includegraphics[width=\linewidth]{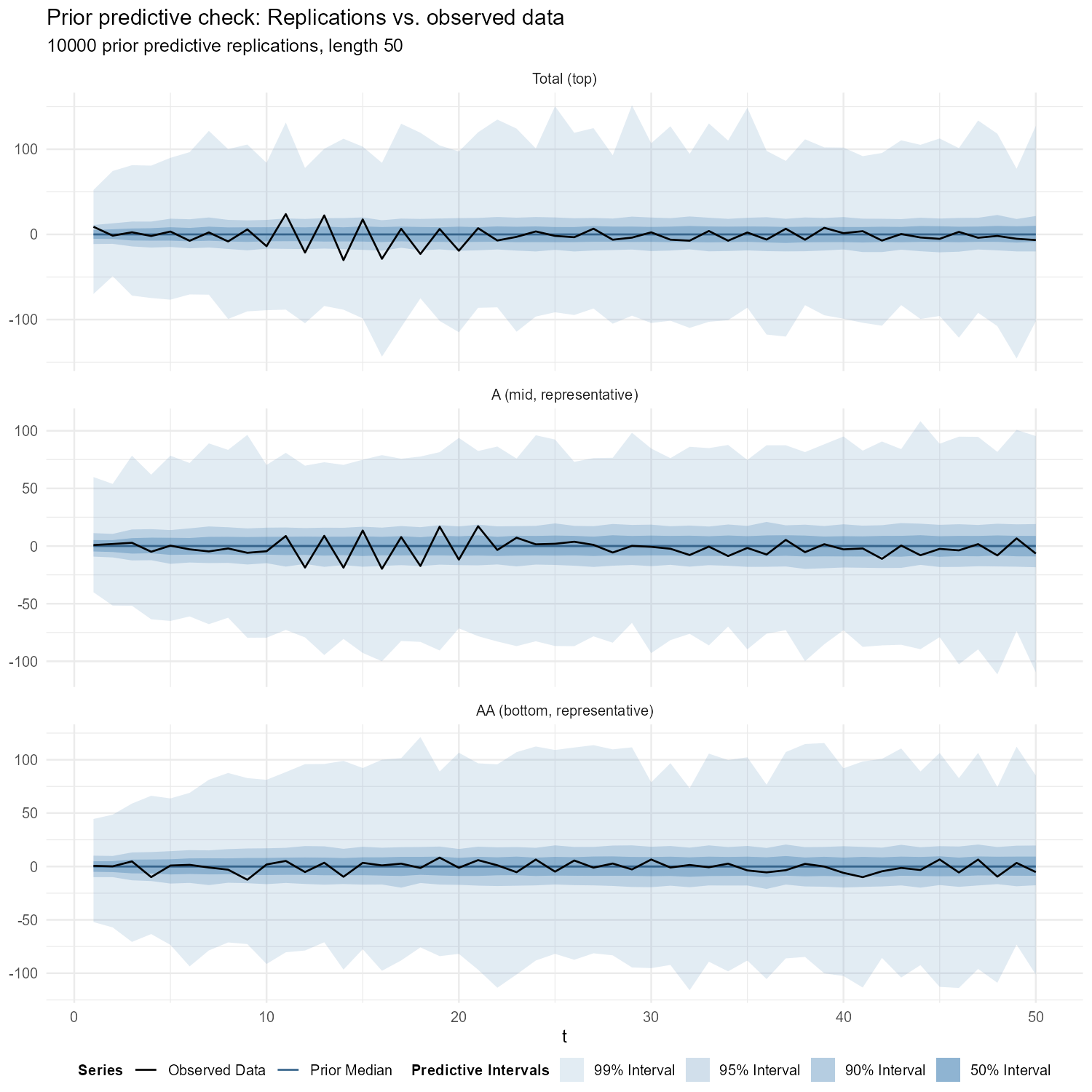}
    \caption{Prior predictive check for the well-specified independent Bayesian model on the simulated data.  }
    \label{fig:prior_predictive_independent}
\end{figure}

\begin{figure}
    \centering
    \includegraphics[width=\linewidth]{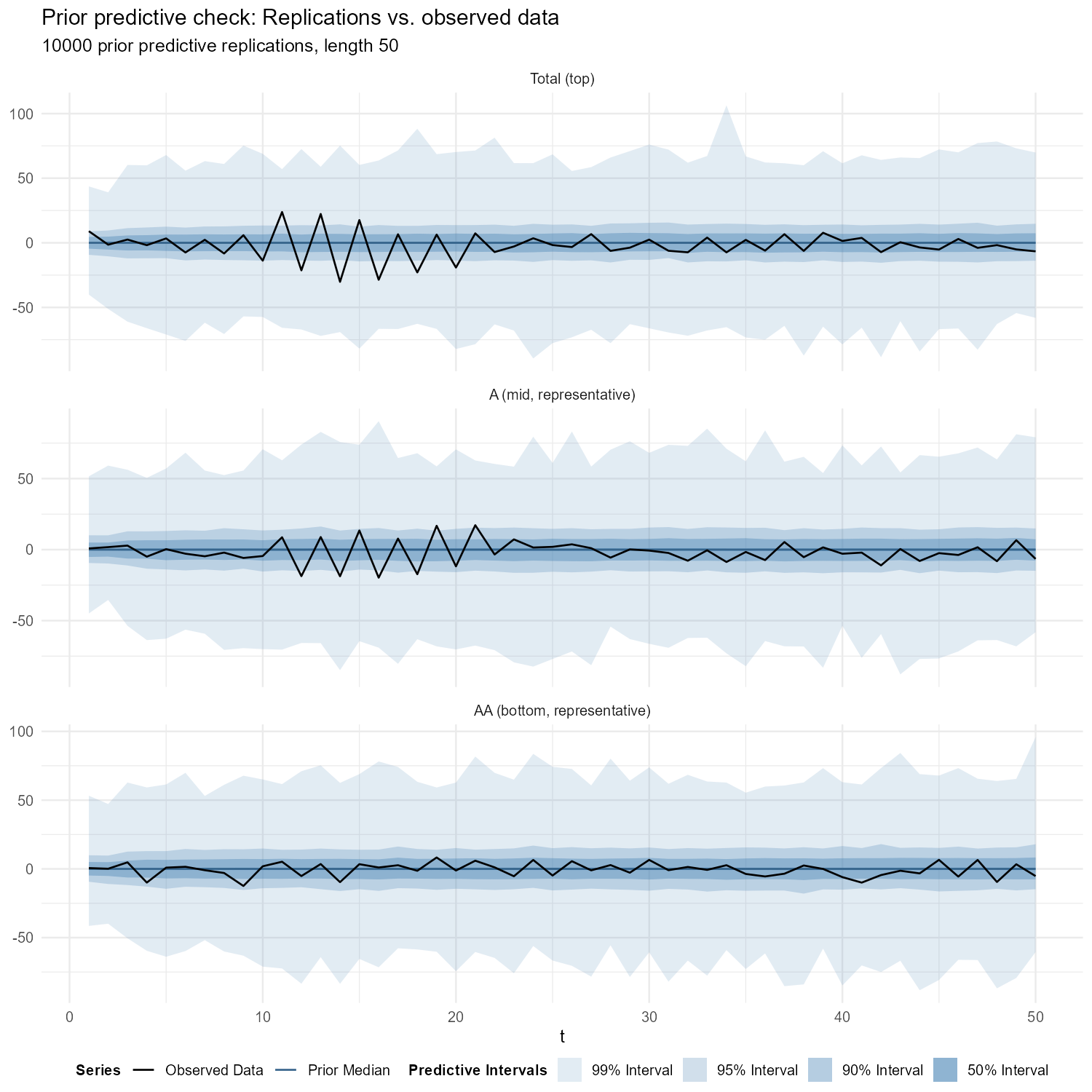}
    \caption{Prior predictive check for the well-specified hierarchical Bayesian model on the simulated data.  }
    \label{fig:prior_predictive_hierarchical}
\end{figure}

\begin{figure}
    \centering
    \includegraphics[width=\linewidth]{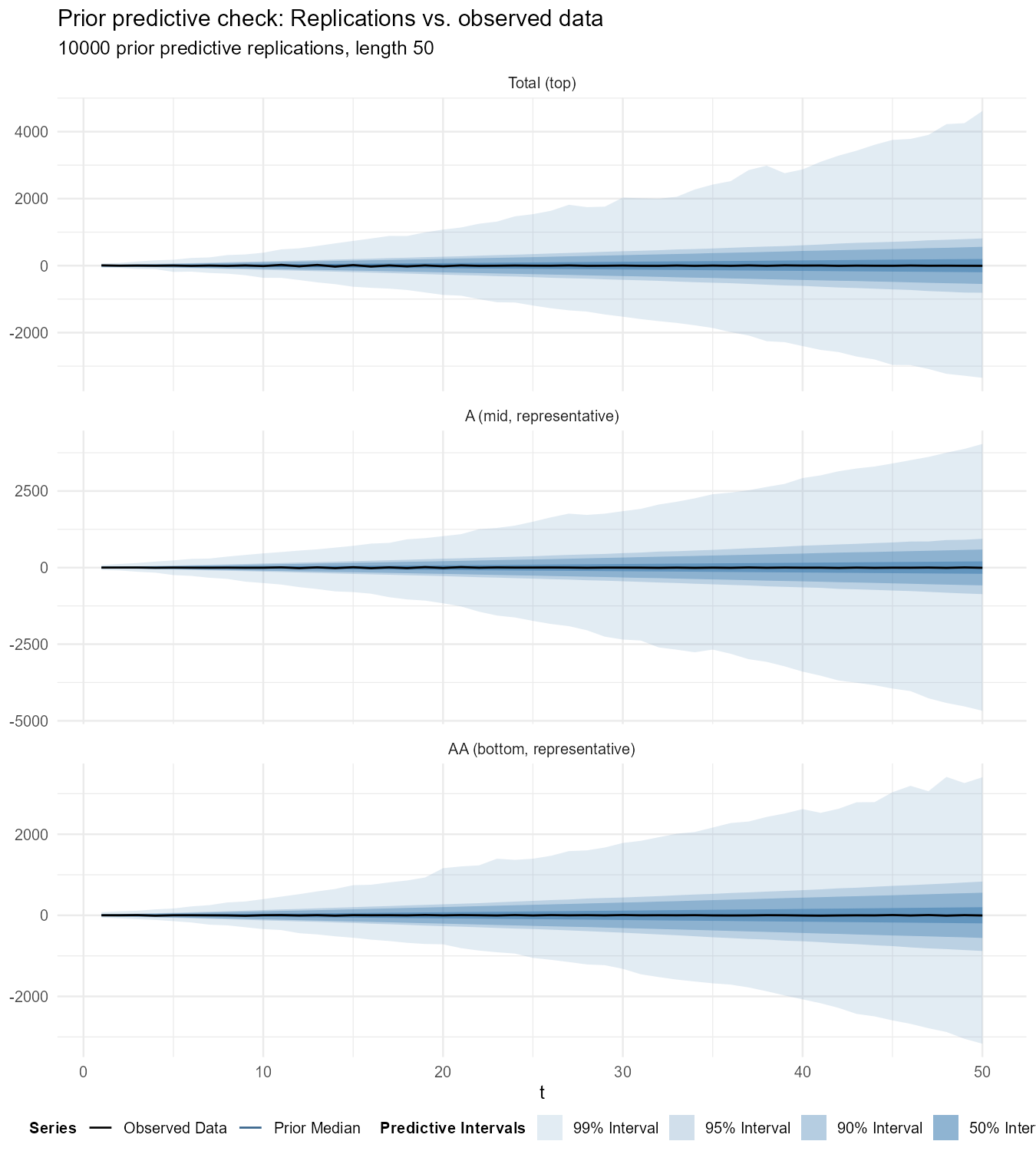}
    \caption{Prior predictive check for the misspecified independent Bayesian model on the simulated data.  }
    \label{fig:prior_predictive_independent_misspecified}
\end{figure}

\begin{figure}
    \centering
    \includegraphics[width=\linewidth]{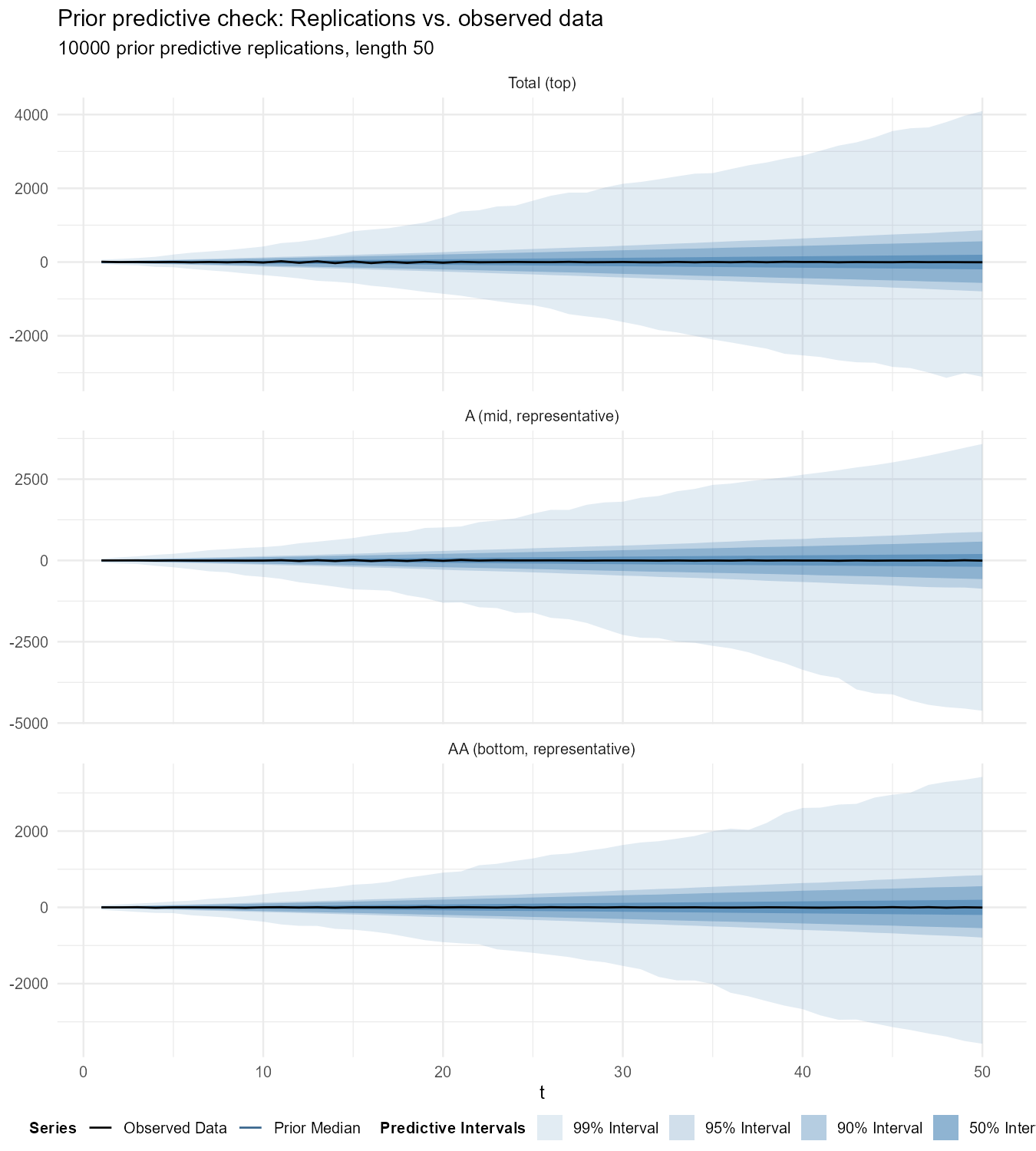}
    \caption{Prior predictive check for the misspecified hierarchical Bayesian model on the simulated data. }
    \label{fig:prior_predictive_hierarchical_misspecified}
\end{figure}

Figures \ref{fig:prior_predictive_independent} to \ref{fig:prior_predictive_hierarchical_misspecified} show the prior predictive checks for the simulation study. We note that the training data falls well within the prior predictive in Figures \ref{fig:prior_predictive_independent} and \ref{fig:prior_predictive_hierarchical}, validating the choice of these priors as appropriately weakly informative in the well-specified case. The base forecasting model is ARMA(2,2), which has stationarity constraints; hence the relatively stable predictive intervals across the range of the data. In Figures \ref{fig:prior_predictive_independent_misspecified} and \ref{fig:prior_predictive_hierarchical_misspecified}, the misspecified ETS model is not stationary and the prior variance is relatively large, hence the variance of the predictive diverges when making predictions over 50 time steps. All data are included well within the prior predictive range. 

\begin{figure}
    \centering
    \includegraphics[width=\linewidth]{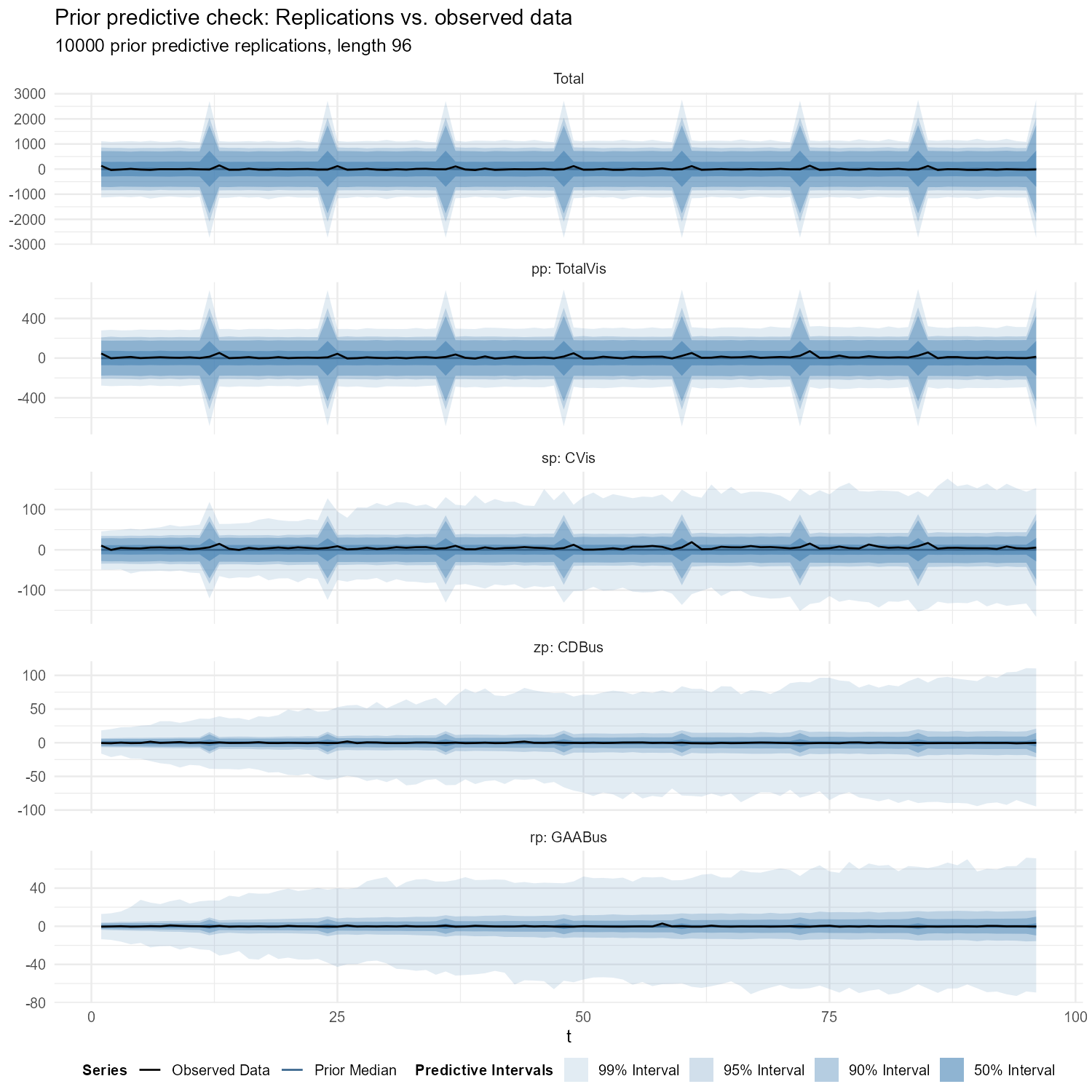}
    \caption{Prior predictive check for the independent Bayesian model on the Australian domestic tourism data. }
    \label{fig:prior_predictive_tourism_independent}
\end{figure}

\begin{figure}
    \centering
    \includegraphics[width=\linewidth]{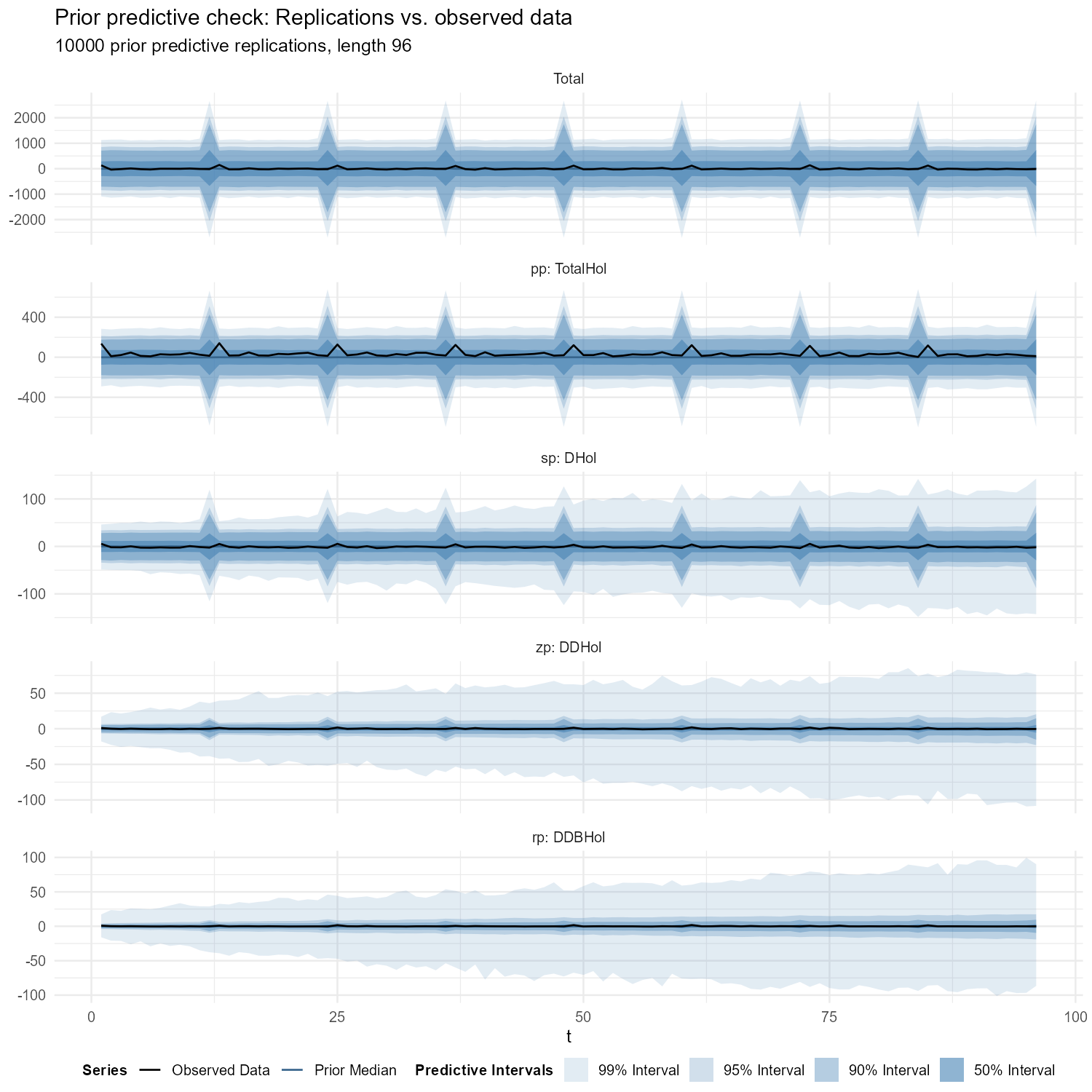}
    \caption{Prior predictive check for the independent Bayesian model on the Australian domestic tourism data. }
    \label{fig:prior_predictive_tourism_hierarchical}
\end{figure}

Figures \ref{fig:prior_predictive_tourism_independent} and \ref{fig:prior_predictive_tourism_hierarchical} show the prior predictive checks for the Australian domestic tourism model. We note the increased variance every 12\textth month, arising from the seasonal sum-to-zero constraint, where the seasonal term for the 12\textth month is determined from the other 11 months and hence is more uncertain, reflected by the increased variance. The first 96 observed data points are well included within the predictive intervals of the prior predictive.

\newpage
\color{black}
\section{Additional data tables for tourism example} \label{appendix:tables_tourism}

\begin{table}[h!]
\begin{center}
\caption{RMSSE averaged overall and RMSE averaged at each level for selected horizons. } \label{tab:overall_RMSE}
\footnotesize
\begin{adjustbox}{max width=\textwidth}
\begin{tabular}{lllllllll}
\multicolumn{9}{l}{\textbf{RMSSE (overall average)}}                                                                                                                                                           \\
Level                       & Method                                                    & $h=1$          & $h=2$          & $h=3$          & $h=6$          & $h=12$         & $h=1-6$   & $h=1-12$  \\ \hline
\rowcolor[HTML]{EFEFEF} 
                            &                                                           &                &                &                &                &                &                &                \\
                            & \multicolumn{1}{l|}{Independent Bayes}                    & 0.856          & 0.858          & 0.858          & 0.863          & 0.876          & 0.860          & 0.865          \\
                            & \multicolumn{1}{l|}{Independent Bayes MinT}               & 0.861          & 0.863          & 0.863          & 0.866          & 0.878          & 0.864          & 0.869          \\
                            & \multicolumn{1}{l|}{Hierarchical Bayes}                   & 0.849          & 0.851          & 0.851          & 0.856          & 0.870          & 0.853          & 0.858          \\
                            & \multicolumn{1}{l|}{Hierarchical Bayes MinT}              & 0.852          & 0.854          & 0.854          & 0.858          & 0.870          & 0.855          & 0.860          \\
                            & \multicolumn{1}{l|}{Hierarchical Bayes softer coherency}  & 0.849          & 0.851          & 0.851          & 0.856          & 0.870          & \textbf{0.852} & 0.858          \\
                            & \multicolumn{1}{l|}{Hierarchical Bayes tighter coherency} & \textbf{0.848} & \textbf{0.850} & 0.851          & 0.856          & 0.870          & \textbf{0.852} & 0.858          \\
                            & \multicolumn{1}{l|}{Hierarchical Bayes total-weighted}    & \textbf{0.848} & \textbf{0.850} & 0.851          & \textbf{0.855} & 0.869          & \textbf{0.852} & \textbf{0.857} \\
\multirow{-8}{*}{Overall}   & \multicolumn{1}{l|}{Hierarchical Bayes purposes-weighted} & \textbf{0.848} & \textbf{0.850} & \textbf{0.850} & \textbf{0.855} & \textbf{0.868} & \textbf{0.852} & \textbf{0.857} \\
                            &                                                           &                &                &                &                &                &                &                \\
\multicolumn{9}{l}{\textbf{RMSE (average at each level)}}                                                                                                                                                      \\
Level                       & Method                                                    & $h=1$          & $h=2$          & $h=3$          & $h=6$          & $h=12$         & $h=1-6$   & $h=1-12$  \\ \hline
\rowcolor[HTML]{EFEFEF} 
                            &                                                           &                &                &                &                &                &                &                \\
                            & \multicolumn{1}{l|}{Independent Bayes}                    & 1670           & 1716           & 1748           & 1824           & 1968           & 1764           & 1817           \\
                            & \multicolumn{1}{l|}{Independent Bayes MinT}               & 1659           & 1689           & 1693           & 1776           & 1925           & 1720           & 1782           \\
                            & \multicolumn{1}{l|}{Hierarchical Bayes}                   & 1653           & 1693           & 1725           & 1808           & 1972           & 1743           & 1806           \\
                            & \multicolumn{1}{l|}{Hierarchical Bayes MinT}              & 1647           & 1675           & \textbf{1688}  & 1776           & 1954           & 1712           & 1782           \\
                            & \multicolumn{1}{l|}{Hierarchical Bayes softer coherency}  & 1686           & 1713           & 1749           & 1852           & 2066           & 1770           & 1861           \\
                            & \multicolumn{1}{l|}{Hierarchical Bayes tighter coherency} & 1704           & 1736           & 1773           & 1876           & 2088           & 1791           & 1884           \\
                            & \multicolumn{1}{l|}{Hierarchical Bayes total-weighted}    & \textbf{1622}  & 1671           & 1706           & \textbf{1773}  & \textbf{1908}  & 1712           & \textbf{1764}  \\
\multirow{-8}{*}{Australia} & \multicolumn{1}{l|}{Hierarchical Bayes purposes-weighted} & 1633           & \textbf{1659}  & 1690           & 1781           & 1970           & \textbf{1709}  & 1785           \\
\rowcolor[HTML]{EFEFEF} 
                            &                                                           &                &                &                &                &                &                &                \\
                            & \multicolumn{1}{l|}{Independent Bayes}                    & 636            & 649            & 652            & 682            & 738            & 659            & 688            \\
                            & \multicolumn{1}{l|}{Independent Bayes MinT}               & 642            & 650            & 651            & 678            & 731            & 658            & 685            \\
                            & \multicolumn{1}{l|}{Hierarchical Bayes}                   & 633            & 645            & 651            & 680            & 738            & 657            & 687            \\
                            & \multicolumn{1}{l|}{Hierarchical Bayes MinT}              & 634            & 644            & 648            & 676            & 733            & 655            & 683            \\
                            & \multicolumn{1}{l|}{Hierarchical Bayes softer coherency}  & 643            & 653            & 664            & 694            & 755            & 669            & 700            \\
                            & \multicolumn{1}{l|}{Hierarchical Bayes tighter coherency} & 650            & 662            & 673            & 702            & 763            & 677            & 707            \\
                            & \multicolumn{1}{l|}{Hierarchical Bayes total-weighted}    & 626            & 642            & 652            & 678            & 731            & 655            & 682            \\
\multirow{-8}{*}{Purposes}  & \multicolumn{1}{l|}{Hierarchical Bayes purposes-weighted} & \textbf{624}   & \textbf{635}   & \textbf{642}   & \textbf{669}   & \textbf{724}   & \textbf{647}   & \textbf{675}   \\
\rowcolor[HTML]{EFEFEF} 
                            &                                                           &                &                &                &                &                &                &                \\
                            & \multicolumn{1}{l|}{Independent Bayes}                    & 178            & 179            & 179            & 181            & 186            & 180            & 182            \\
                            & \multicolumn{1}{l|}{Independent Bayes MinT}               & 177            & 178            & 178            & 179            & 185            & 178            & 180            \\
                            & \multicolumn{1}{l|}{Hierarchical Bayes}                   & 176            & 177            & 178            & 180            & 186            & 178            & 181            \\
                            & \multicolumn{1}{l|}{Hierarchical Bayes MinT}              & \textbf{175}   & \textbf{176}   & \textbf{177}   & \textbf{178}   & \textbf{184}   & \textbf{177}   & \textbf{179}   \\
                            & \multicolumn{1}{l|}{Hierarchical Bayes softer coherency}  & 176            & 177            & 178            & 180            & 186            & 178            & 181            \\
                            & \multicolumn{1}{l|}{Hierarchical Bayes tighter coherency} & 176            & 177            & 179            & 181            & 186            & 179            & 181            \\
                            & \multicolumn{1}{l|}{Hierarchical Bayes total-weighted}    & 176            & 177            & 178            & 180            & 185            & 178            & 180            \\
\multirow{-8}{*}{States}    & \multicolumn{1}{l|}{Hierarchical Bayes purposes-weighted} & \textbf{175}   & 177            & \textbf{177}   & 179            & \textbf{184}   & 178            & 180            \\
\rowcolor[HTML]{EFEFEF} 
                            &                                                           &                &                &                &                &                &                &                \\
                            & \multicolumn{1}{l|}{Independent Bayes}                    & 78.9           & 79.1           & 79.1           & 79.5           & 80.8           & 79.3           & 79.8           \\
                            & \multicolumn{1}{l|}{Independent Bayes MinT}               & 79.1           & 79.2           & 79.2           & 79.4           & 80.7           & 79.3           & 79.8           \\
                            & \multicolumn{1}{l|}{Hierarchical Bayes}                   & 78.2           & 78.4           & 78.5           & 78.9           & 80.3           & 78.6           & 79.1           \\
                            & \multicolumn{1}{l|}{Hierarchical Bayes MinT}              & 78.3           & 78.4           & 78.5           & 78.8           & \textbf{80.0}  & 78.6           & 79.0           \\
                            & \multicolumn{1}{l|}{Hierarchical Bayes softer coherency}  & \textbf{78.1}  & \textbf{78.3}  & 78.5           & 78.8           & 80.2           & \textbf{78.5}  & 79.0           \\
                            & \multicolumn{1}{l|}{Hierarchical Bayes tighter coherency} & \textbf{78.1}  & \textbf{78.3}  & \textbf{78.4}  & 78.8           & 80.2           & \textbf{78.5}  & 79.0           \\
                            & \multicolumn{1}{l|}{Hierarchical Bayes total-weighted}    & \textbf{78.1}  & \textbf{78.3}  & 78.5           & \textbf{78.7}  & \textbf{80.0}  & \textbf{78.5}  & 79.0           \\
\multirow{-8}{*}{Zones}     & \multicolumn{1}{l|}{Hierarchical Bayes purposes-weighted} & 78.2           & \textbf{78.3}  & 78.5           & \textbf{78.7}  & \textbf{80.0}  & \textbf{78.5}  & \textbf{78.9}  \\
\rowcolor[HTML]{EFEFEF} 
                            &                                                           &                &                &                &                &                &                &                \\
                            & \multicolumn{1}{l|}{Independent Bayes}                    & 40.9           & 40.8           & 40.9           & 41.0           & 41.5           & 40.9           & 41.1           \\
                            & \multicolumn{1}{l|}{Independent Bayes MinT}               & 41.1           & 41.1           & 41.1           & 41.2           & 41.6           & 41.1           & 41.3           \\
                            & \multicolumn{1}{l|}{Hierarchical Bayes}                   & \textbf{40.5}  & 40.5           & 40.6           & 40.7           & 41.2           & 40.6           & 40.8           \\
                            & \multicolumn{1}{l|}{Hierarchical Bayes MinT}              & 40.7           & 40.6           & 40.7           & 40.8           & 41.2           & 40.7           & 40.9           \\
                            & \multicolumn{1}{l|}{Hierarchical Bayes softer coherency}  & \textbf{40.5}  & 40.5           & \textbf{40.5}  & 40.7           & 41.2           & 40.6           & \textbf{40.7}  \\
                            & \multicolumn{1}{l|}{Hierarchical Bayes tighter coherency} & \textbf{40.5}  & \textbf{40.4}  & \textbf{40.5}  & \textbf{40.6}  & 41.2           & \textbf{40.5}  & \textbf{40.7}  \\
                            & \multicolumn{1}{l|}{Hierarchical Bayes total-weighted}    & \textbf{40.5}  & 40.5           & \textbf{40.5}  & \textbf{40.6}  & 41.2           & 40.6           & \textbf{40.7}  \\
\multirow{-8}{*}{Regions}   & \multicolumn{1}{l|}{Hierarchical Bayes purposes-weighted} & \textbf{40.5}  & 40.5           & \textbf{40.5}  & \textbf{40.6}  & \textbf{41.1}  & 40.6           & \textbf{40.7} 
\end{tabular}
\end{adjustbox}
\end{center}
\end{table}

\begin{table}[h]
\begin{center}
\caption{ES and CRPS for selected horizons and averages at all levels. } \label{tab:overall_CRPS}
\footnotesize
\begin{adjustbox}{max width=\textwidth}
\begin{tabular}{lllllllll}
\multicolumn{9}{l}{\textbf{ES (overall)}}                                                                                                                                                               \\
Level                       & Method                                                    & $h=1$         & $h=2$         & $h=3$         & $h=6$         & $h=12$        & $h=1-6$  & $h=1-12$ \\ \hline
\rowcolor[HTML]{EFEFEF} 
                            &                                                           &               &               &               &               &               &               &               \\
                            & \multicolumn{1}{l|}{Independent Bayes}                    & 1878          & 1903          & 1922          & 1975          & 2095          & 1933          & 1985          \\
                            & \multicolumn{1}{l|}{Independent Bayes MinT}               & 1867          & 1888          & 1891          & 1946          & \textbf{2050} & 1907          & 1957          \\
                            & \multicolumn{1}{l|}{Hierarchical Bayes}                   & 1859          & 1884          & 1905          & 1963          & 2090          & 1916          & 1971          \\
                            & \multicolumn{1}{l|}{Hierarchical Bayes MinT}              & 1849          & 1869          & 1881          & 1943          & 2078          & 1897          & 1953          \\
                            & \multicolumn{1}{l|}{Hierarchical Bayes softer coherency}  & 1876          & 1895          & 1924          & 1986          & 2127          & 1933          & 1998          \\
                            & \multicolumn{1}{l|}{Hierarchical Bayes tighter coherency} & 1887          & 1910          & 1934          & 2001          & 2149          & 1945          & 2012          \\
                            & \multicolumn{1}{l|}{Hierarchical Bayes total-weighted}    & \textbf{1832} & 1867          & 1894          & 1945          & 2056          & 1897          & \textbf{1947} \\
\multirow{-8}{*}{Overall}   & \multicolumn{1}{l|}{Hierarchical Bayes purposes-weighted} & 1834          & \textbf{1859} & \textbf{1881} & \textbf{1935} & 2068          & \textbf{1889} & \textbf{1947} \\
                            &                                                           &               &               &               &               &               &               &               \\
\multicolumn{9}{l}{\textbf{CRPS (average at each level)}}                                                                                                                                               \\
Level                       & Method                                                    & $h=1$         & $h=2$         & $h=3$         & $h=6$         & $h=12$        & $h=1-6$  & $h=1-12$ \\ \hline
\rowcolor[HTML]{EFEFEF} 
                            &                                                           &               &               &               &               &               &               &               \\
                            & \multicolumn{1}{l|}{Independent Bayes}                    & 946           & 974           & 993           & 1037          & 1133          & 1001          & 1036          \\
                            & \multicolumn{1}{l|}{Independent Bayes MinT}               & 957           & 978           & 976           & 1027          & 1126          & 994           & 1034          \\
                            & \multicolumn{1}{l|}{Hierarchical Bayes}                   & 935           & 960           & 978           & 1026          & 1137          & 988           & 1030          \\
                            & \multicolumn{1}{l|}{Hierarchical Bayes MinT}              & 961           & 982           & 988           & 1044          & 1178          & 1004          & 1054          \\
                            & \multicolumn{1}{l|}{Hierarchical Bayes softer coherency}  & 960           & 978           & 1000          & 1062          & 1205          & 1012          & 1072          \\
                            & \multicolumn{1}{l|}{Hierarchical Bayes tighter coherency} & 972           & 993           & 1014          & 1078          & 1222          & 1026          & 1088          \\
                            & \multicolumn{1}{l|}{Hierarchical Bayes total-weighted}    & \textbf{917}  & 948           & 968           & \textbf{1012} & \textbf{1109} & \textbf{972}  & \textbf{1010} \\
\multirow{-8}{*}{Australia} & \multicolumn{1}{l|}{Hierarchical Bayes purposes-weighted} & 927           & \textbf{942}  & \textbf{960}  & 1014          & 1145          & \textbf{972}  & 1023          \\
\rowcolor[HTML]{EFEFEF} 
                            &                                                           &               &               &               &               &               &               &               \\
                            & \multicolumn{1}{l|}{Independent Bayes}                    & 351           & 357           & 360           & 376           & 410           & 364           & 381           \\
                            & \multicolumn{1}{l|}{Independent Bayes MinT}               & 355           & 360           & 361           & 376           & 409           & 365           & 381           \\
                            & \multicolumn{1}{l|}{Hierarchical Bayes}                   & 349           & 355           & 359           & 375           & 410           & 362           & 379           \\
                            & \multicolumn{1}{l|}{Hierarchical Bayes MinT}              & 352           & 357           & 360           & 377           & 414           & 364           & 381           \\
                            & \multicolumn{1}{l|}{Hierarchical Bayes softer coherency}  & 354           & 359           & 365           & 383           & 420           & 368           & 386           \\
                            & \multicolumn{1}{l|}{Hierarchical Bayes tighter coherency} & 358           & 364           & 370           & 388           & 425           & 373           & 391           \\
                            & \multicolumn{1}{l|}{Hierarchical Bayes total-weighted}    & 344           & 352           & 358           & 373           & 405           & 359           & 376           \\
\multirow{-8}{*}{Purposes}  & \multicolumn{1}{l|}{Hierarchical Bayes purposes-weighted} & \textbf{343}  & \textbf{348}  & \textbf{353}  & \textbf{368}  & \textbf{401}  & \textbf{355}  & \textbf{372}  \\
\rowcolor[HTML]{EFEFEF} 
                            &                                                           &               &               &               &               &               &               &               \\
                            & \multicolumn{1}{l|}{Independent Bayes}                    & 95.3          & 95.8          & 96.3          & 97.7          & 101           & 96.6          & 98.0          \\
                            & \multicolumn{1}{l|}{Independent Bayes MinT}               & 95.3          & 95.9          & 96.0          & 97.0          & 100           & 96.3          & 97.6          \\
                            & \multicolumn{1}{l|}{Hierarchical Bayes}                   & 94.5          & 95.1          & 95.7          & 97.0          & 100           & 95.9          & 97.4          \\
                            & \multicolumn{1}{l|}{Hierarchical Bayes MinT}              & \textbf{93.9} & \textbf{94.6} & \textbf{95.0} & \textbf{95.9} & \textbf{99.2} & \textbf{95.2} & \textbf{96.4} \\
                            & \multicolumn{1}{l|}{Hierarchical Bayes softer coherency}  & 94.5          & 95.0          & 95.7          & 97.0          & 100           & 95.9          & 97.4          \\
                            & \multicolumn{1}{l|}{Hierarchical Bayes tighter coherency} & 94.6          & 95.2          & 95.8          & 97.2          & 101           & 96.0          & 97.5          \\
                            & \multicolumn{1}{l|}{Hierarchical Bayes total-weighted}    & 94.2          & 94.9          & 95.7          & 96.8          & 99.7          & 95.8          & 97.2          \\
\multirow{-8}{*}{States}    & \multicolumn{1}{l|}{Hierarchical Bayes purposes-weighted} & 94.0          & \textbf{94.6} & 95.2          & 96.3          & 99.5          & 95.4          & 96.7          \\
\rowcolor[HTML]{EFEFEF} 
                            &                                                           &               &               &               &               &               &               &               \\
                            & \multicolumn{1}{l|}{Independent Bayes}                    & 41.3          & 41.4          & 41.5          & 41.7          & 42.5          & 41.6          & 41.9          \\
                            & \multicolumn{1}{l|}{Independent Bayes MinT}               & 41.7          & 41.8          & 41.9          & 42.1          & 42.9          & 41.9          & 42.2          \\
                            & \multicolumn{1}{l|}{Hierarchical Bayes}                   & 41.0          & 41.1          & 41.2          & 41.4          & 42.2          & 41.2          & 41.5          \\
                            & \multicolumn{1}{l|}{Hierarchical Bayes MinT}              & \textbf{40.7} & \textbf{40.8} & \textbf{40.8} & \textbf{41.0} & \textbf{41.7} & \textbf{40.9} & \textbf{41.1} \\
                            & \multicolumn{1}{l|}{Hierarchical Bayes softer coherency}  & 40.9          & 41.0          & 41.1          & 41.4          & 42.1          & 41.2          & 41.5          \\
                            & \multicolumn{1}{l|}{Hierarchical Bayes tighter coherency} & 40.9          & 41.0          & 41.1          & 41.3          & 42.1          & 41.1          & 41.4          \\
                            & \multicolumn{1}{l|}{Hierarchical Bayes total-weighted}    & 40.9          & 41.0          & 41.1          & 41.3          & 42.1          & 41.1          & 41.4          \\
\multirow{-8}{*}{Zones}     & \multicolumn{1}{l|}{Hierarchical Bayes purposes-weighted} & 40.9          & 41.0          & 41.1          & 41.3          & 42.0          & 41.1          & 41.4          \\
\rowcolor[HTML]{EFEFEF} 
                            &                                                           &               &               &               &               &               &               &               \\
                            & \multicolumn{1}{l|}{Independent Bayes}                    & 21.0          & 21.0          & 21.0          & 21.2          & 21.4          & 21.1          & 21.2          \\
                            & \multicolumn{1}{l|}{Independent Bayes MinT}               & 21.7          & 21.7          & 21.8          & 21.9          & 22.2          & 21.8          & 21.9          \\
                            & \multicolumn{1}{l|}{Hierarchical Bayes}                   & \textbf{20.8} & \textbf{20.8} & 20.9          & 21.0          & 21.3          & 20.9          & 21.0          \\
                            & \multicolumn{1}{l|}{Hierarchical Bayes MinT}              & 21.0          & 21.0          & 21.0          & 21.1          & 21.4          & 21.0          & 21.1          \\
                            & \multicolumn{1}{l|}{Hierarchical Bayes softer coherency}  & \textbf{20.8} & \textbf{20.8} & \textbf{20.8} & \textbf{20.9} & \textbf{21.2} & 20.9          & 21.0          \\
                            & \multicolumn{1}{l|}{Hierarchical Bayes tighter coherency} & \textbf{20.8} & \textbf{20.8} & \textbf{20.8} & \textbf{20.9} & \textbf{21.2} & \textbf{20.8} & \textbf{20.9} \\
                            & \multicolumn{1}{l|}{Hierarchical Bayes total-weighted}    & \textbf{20.8} & \textbf{20.8} & \textbf{20.8} & \textbf{20.9} & \textbf{21.2} & 20.9          & 21.0          \\
\multirow{-8}{*}{Regions}   & \multicolumn{1}{l|}{Hierarchical Bayes purposes-weighted} & \textbf{20.8} & \textbf{20.8} & \textbf{20.8} & \textbf{20.9} & \textbf{21.2} & 20.9          & 21.0         
\end{tabular}
\end{adjustbox}
\end{center}
\end{table}

\end{document}